\documentclass[12pt]{article}
\pdfoutput=1

\usepackage[utf8]{inputenc}
\usepackage[left=2.55cm, right=2.55cm, top=2.55cm, bottom=2.55cm]{geometry}
\usepackage{amsmath,amssymb,amsbsy}
\usepackage{slashed}
\usepackage{xcolor}
\usepackage{graphicx}
\usepackage{url}
\usepackage{cancel}
\usepackage{cite}
\usepackage[colorlinks=true,allcolors=blue,pdfborder={0 0 0},linktocpage=false,pdfencoding=auto]{hyperref}
\usepackage{tabularx,booktabs}
\usepackage{multicol}
\usepackage{feynmp}
\usepackage{units}
\usepackage{xspace}
\usepackage[labelfont=bf]{caption}
\usepackage[section]{placeins}
\usepackage{subcaption}
\usepackage{soul}
\usepackage{bbold}
\usepackage{parskip}
\usepackage{float}
\usepackage{tabulary}

\definecolor{darkred}{rgb}{0.6,0,0}
\definecolor{darkpurple}{rgb}{0.5,0,0.5}

\newcommand{\code}[1]{\texttt{#1}}
\newcommand{\beqn}{\begin{eqnarray}}
\newcommand{\eeqn}{\end{eqnarray}}

\begin{document}

\author{Amin Aboubrahim$^a$\footnote{\href{mailto:aabouibr@uni-muenster.de}{aabouibr@uni-muenster.de}}~, Michael Klasen$^a$\footnote{\href{mailto:michael.klasen@uni-muenster.de}{michael.klasen@uni-muenster.de}}~~and Pran Nath$^b$\footnote{\href{mailto:p.nath@northeastern.edu}{p.nath@northeastern.edu}} \\~\\
$^{a}$\textit{\normalsize Institut f\"ur Theoretische Physik, Westf\"alische Wilhelms-Universit\"at M\"unster,} \\
\textit{\normalsize Wilhelm-Klemm-Stra{\ss}e 9, 48149 M\"unster, Germany} \\
$^{b}$\textit{\normalsize Department of Physics, Northeastern University, Boston, MA 02115-5000, USA} \\
}

\title{\vspace{-2cm}\begin{flushright}
{\small MS-TP-22-03}
\end{flushright}
\vspace{1cm}
\Large \bf
Analyzing the Hubble tension through hidden sector dynamics in the early universe
 \vspace{0.5cm}}

\date{}
\maketitle

\begin{abstract}

The recent analysis from the SH0ES Collaboration has confirmed the existence of a Hubble 
tension between measurements at high redshift  ($z> 1000$) and  at low redshift ($z<1$)
at the $5\sigma$ level with the low redshift measurement giving a higher value. 
In this work we propose a particle physics model that can help alleviate the Hubble tension 
 via an out-of-equilibrium hidden sector coupled to the visible sector.
The particles that populate the dark sector consist of a dark fermion, which acts as dark matter, 
 a dark photon, a massive scalar and a massless pseudo-scalar. Assuming no initial population of
 particles in the dark sector, feeble couplings between the visible and the hidden sectors via
 kinetic mixing  populate
 the dark sector even though the number densities of hidden sector particles 
 never reach their equilibrium distribution and the two sectors remain at different temperatures. 
 A cosmologically consistent analysis is  presented where a correlated evolution of the visible and the hidden sectors with coupled Boltzmann equations involving two temperatures, one for the visible
 sector and the other for the hidden sector, is carried out. The relic density of the dark matter constituted of dark fermions
is computed in this two-temperature formalism.  
  As a consequence, BBN predictions are upheld with a minimal contribution to $\Delta N_{\rm eff}$.  However, the out-of-equilibrium decay of the massive scalar to the massless pseudo-scalar close to the recombination time causes an increase in $\Delta N_{\rm eff}$ that can help weaken the
  Hubble tension.

\end{abstract}

\numberwithin{equation}{section}

\newpage

{  \hrule height 0.4mm \hypersetup{colorlinks=black,linktocpage=true} \tableofcontents
\vspace{0.5cm}
 \hrule height 0.4mm}

\section{Introduction}

Recently the SH0ES Collaboration~\cite{Riess:2021jrx} has confirmed at $5\sigma$ level 
the tension between measurements of the Hubble parameter $H_0$ at high redshift ($z>1000$) and at
low redshift ($z<1$). Based on the Lambda-cold dark matter ($\Lambda$CDM)
 model and an analysis of data from the times of Big Bang Nucleosynthesis (BBN), the Cosmic Microwave Background (CMB) and Baryon Acoustic Oscillations (BAO), the Hubble parameter at high $z$ is determined to be~\cite{Planck:2018vyg}
\begin{align}
H_0 = (67.4\pm 0.5)\text{ km/s}/\text{Mpc},
\end{align}
while the low $z$ value arising from the analysis of Cepheids and SNIa gives~\cite{Riess:2021jrx}
\begin{align}
H_0 = (73.04\pm 1.04)\text{ km/s}/\text{Mpc},
\end{align}
which leads to a $5\sigma$ difference between the Planck result and the result arising from local observation.  Since the Planck analysis is based on the $\Lambda$CDM model, the discrepancy 
between the high $z$ and the low $z$ results  
represents an indication of physics beyond the standard  $\Lambda$CDM model.
There is a large number of theory models which attempt to explain this difference (for a review see ref.~\cite{DiValentino:2021izs} and references therein, e.g., ref.~\cite{Solomon:2022qqf}).
 One of the simplest avenues is the possibility of having extra relativistic degrees of freedom present during recombination,
 which would cause an increase in $H_0$ and bring it closer to its measured local value. Extensions of the SM with new particles in thermal equilibrium with the neutrinos, such as a majoron~\cite{Fernandez-Martinez:2021ypo,Escudero:2021rfi}, or with an extra $U(1)$ symmetry and $Z'$ bosons decaying to neutrinos or otherwise~\cite{Gehrlein:2019iwl,Escudero:2019gzq} can explain the tension. We note here that  models that can successfully explain the 
Hubble tension are those where the extra degrees of freedom would affect only the dynamics in the CMB era, but not impact BBN. This is so because the precise predictions of light element synthesis during BBN can be translated into strong constraints on the effective degrees of freedom such that
 $N_{\rm eff}^{\rm BBN}=2.88\pm 0.27$ at 68\% C.L.~\cite{Cyburt:2015mya,Pitrou:2018cgg}. This appears to be consistent with the SM prediction of $N_{\rm eff}^{\rm SM}\simeq 3.046$~\cite{Mangano:2005cc}.
   Since the effective degrees of freedom are tied to the energy density in the universe, which affects the Hubble parameter, any deviation from $N_{\rm eff}$ at BBN would have an impact on the expansion rate of the universe as well as on predictions of light element synthesis during this epoch. Thus any model looking to add extra relativistic degrees freedom at CMB time to help alleviate the Hubble tension should not violate the BBN constraints on $N_{\rm eff}$.  
 
Further, $N_{\rm eff}$ is also constrained, but to a lesser extent, during the period the CMB was generated. This is so because extra relativistic degrees of freedom affect the CMB power spectrum, which has been measured using Planck polarization data and BAO, so that~\cite{Planck:2018vyg}
\begin{equation}
N_{\rm eff}=2.99^{+0.34}_{-0.33}\quad (95\%~\text{C.L., TT,TE,EE+lowE+lensing+BAO)}. 
\end{equation}
Two earlier local measurements of $H_0$~\cite{Riess:2018uxu,Riess:2019cxk} (dubbed R18 and R19) have shown an increasing discrepancy with high redshift values of $H_0$ at 3.7$\sigma$ and $4.4\sigma$ deviation, respectively. The preferred $N_{\rm eff}$ value, when combining the local measurements of $H_0$ with Planck and BAO data, becomes~\cite{Planck:2018vyg}
\begin{equation}
N_{\rm eff}=3.27\pm 0.15\quad (68\%~\text{C.L., TT,TE,EE+lowE+lensing+BAO+R18)}. 
\end{equation}
Extra relativistic degrees of freedom, $\Delta N_{\rm eff}$, increase the sound horizon $r_*$ during recombination. Since the acoustic scale $\theta_*=r_*/D_M$
is precisely measured\footnote{$D_M=\int_0^{z_*} dz/H(z)$, where $z_*$ is the redshift at the time of last scattering.}, then $D_M\propto 1/H_0$ must decrease, or $H_0$ must increase to keep inline with experimental measurements. Based on the difference between the early and local measurements of $H_0$, it  was suggested that
\begin{equation}
0.2 \lesssim\Delta N_{\rm eff}\lesssim 0.5\quad \text{(CMB+BAO+R18),~\cite{Planck:2018vyg}}
\end{equation}  
and
 \begin{equation}
0.2 \lesssim\Delta N_{\rm eff}\lesssim 0.4\quad (\text{CMB+BAO+Pantheon~\cite{Pan-STARRS1:2017jku}+R19+BBN),~\cite{Seto:2021xua,Seto:2021tad,Vagnozzi:2019ezj}}
\end{equation}
could relax the Hubble tension between high redshift values of $H_0$ and the local R18 and R19 measurements\footnote{For an estimate of the Hubble parameter in different cosmologies using the Pantheon data, see refs.~\cite{Dainotti:2021pqg,Dainotti:2022bzg}.}. In light of the most recent measurement of $H_0$ by the SH0ES Collaboration~\cite{Riess:2021jrx} one can expect a revised range in order to accommodate higher values of $\Delta N_{\rm eff}$. 

In this work we present a cosmologically consistent model based on the 
 Stueckelberg extension of the SM with a hidden sector to help alleviate
  the Hubble tension between the high and the low  $z$ measurements. By a cosmologically consistent model we mean that 
 we take into account the fact that the hidden sector and the visible sector in general live in 
 different heat baths at different temperatures and 
 the Boltzmann equations that evolve the number density of the particle species in the visible 
 and in the hidden sectors must use a two-temperature evolution which, however, are 
 correlated because of the couplings between the two sectors.
  Thus the hidden sector possesses a $U(1)_X$ gauge symmetry and the couplings
  between the hidden $U(1)_X$ and the hypercharge $U(1)_Y$ can occur via kinetic 
  mixing and Stueckelberg mass mixing. We assume that, aside from the $U(1)_X$ gauge field (which gives the dark photon $\gamma'$), the hidden sector contains a Dirac fermion $D$, a scalar field $S$ and a pseudo-scalar field $\phi$. The coupling between the visible
  sector and the hidden sector is assumed feeble while that between the dark fermion and the
  dark photon could be of normal size, i.e., of size $\mathcal{O}(1)$. Assuming that there was no initial particle density of the hidden sector particles, the feeble interactions due to small kinetic 
 mixing between the hidden sector $U(1)_X$ and the visible sector $U(1)_Y$ gauge groups allow for a gradual buildup of particle density in the hidden sector. Dirac fermions $D$ (which act as dark matter) and  dark photons $\gamma'$ can be generated from the SM via $2\to 2$ and $2\to 1$ processes. The  $U(1)_X$ gauge coupling $g_X$ is assumed strong enough to allow $D$ and $\gamma'$ to enter into chemical equilibrium in the dark sector.   
It is to be noted that  
   just like the SM thermal bath of massless photons, the dark sector can generate its own thermal bath which we assume here to consist of massless pseudoscalar particle $\phi$. 
   Along with $\phi$, a light long-lived scalar particle $S$ can be produced. Both $\phi$ and $S$ have no direct couplings with the SM and are produced purely through dark sector interactions, which we assume to be at a different temperature than for the SM. The resolution of the Hubble tension comes from the decay $S\to\phi\phi$ prior to recombination, which adds to the relativistic degrees of freedom at this epoch and drive the Hubble parameter to larger values. Particle physics models that attempt to explain the Hubble tension either assume the visible and the hidden sectors are initially in thermal contact and decouple later or they start as out-of-equilibrium systems and eventually a hidden sector particle thermalizes with the neutrino sector before decaying close to recombination. Our analysis does not involve the neutrinos and the resolution of the Hubble tension comes entirely from the dark sector. Further, the model discussed here can explain the Hubble tension while preserving BBN predictions due to the fact that the particles $S$ and $\phi$ do not trace their equilibrium distributions in the dark sector owing to the small coupling between them and $D$ and $\gamma'$. The model parameters are constrained by the DM relic density, limits on the dark photon mass, and by the chemical and kinetic decoupling of the species and $\Delta N_{\rm eff}$ at BBN. 

The paper is organized as follows. In section~\ref{sec:model} we give an overview of the model used in the analysis. In section~\ref{sec:temp} we discuss the set of coupled Boltzmann equations used to evolve the particle yields and the temperatures of the visible and hidden sectors. A discussion of kinetic decoupling of hidden sector particles is given in section~\ref{sec:keq}. In section~\ref{sec:numerical} we give a numerical analysis of the particle yields and the temperature evolution of the sectors taking into accounts the relevant experimental constraints. A discussion on the different contributions to $\Delta N_{\rm eff}$ is given in section~\ref{sec:neff} followed by an analysis of energy densities and $\Delta N_{\rm eff}$ up to recombination in section~\ref{sec:reheating}. Conclusions are given section~\ref{sec:conclu}. Further detail of the analysis is given in section~\ref{sec:append}. 

\section{The model}\label{sec:model}

We now briefly describe the hidden sector model and its coupling to the visible sector which
is used in the analysis of this work. The model is a $U(1)_X$ extension of the SM gauge group which has gauge kinetic mixing~\cite{Holdom:1985ag,Holdom:1990xp} between the two abelian gauge groups. The new gauge field obtains its mass through the Stueckelberg mechanism~\cite{Kors:2004dx,Kors:2004ri} 
 for the gauge fields. Thus the extended model has the Lagrangian 
  \begin{equation}
\mathcal{L}=\mathcal{L}_{\rm SM}+\Delta\mathcal{L},
\label{totL}
\end{equation}
where $\mathcal{L}_{\rm SM}$ is the Standard Model Lagrangian, and  
\begin{align}
\label{totL1}
\Delta\mathcal{L} =&-\frac{1}{4}C_{\mu\nu}C^{\mu\nu}+i\bar D \gamma^\mu \partial_\mu D -m_D \bar D D - \frac{1}{2} 
(\partial_\mu \phi \partial^\mu \phi)- \frac{1}{2} 
(\partial_\mu S \partial^\mu S) \nonumber\\
&+g_X Q_X \bar D \gamma^\mu D C_\mu + y_\phi \bar{D}\gamma_5 D\phi+y_S D\bar{D} S \nonumber \\
&-\frac{\delta}{2}C_{\mu\nu}B^{\mu\nu}-\frac{1}{2}(\partial_{\mu}\sigma+M_1 C_{\mu}+M_2 B_{\mu})^2,
\end{align}
contains additional terms arising from the hidden sector and its coupling to the visible sector
through kinetic energy mixing and Stueckelberg mass mixing. In Eq.~(\ref{totL1}), $C_\mu$ is the gauge field
of the $U(1)_X$ and $B_\mu$ is the gauge field
of the $U(1)_Y$ and $C_{\mu\nu}= \partial_\mu C_\nu -\partial_\nu C_\mu$ and $B_{\mu\nu}=
\partial_\mu B_\nu- \partial_\nu B_\mu$.
The axionic field $\sigma$ transforms dually under $U(1)_X$ and $U(1)_Y$ 
so that the mass term remains gauge invariant and can be discarded in the unitary gauge.
The dark fermions, with a charge $Q_X$ under $U(1)_X$, interact with the gauge field $C_\mu$ via the 
gauge interaction with strength $g_X$,
 and with the (pseudo-)scalar field ($\phi$) $S$ with strength ($y_\phi$) $y_S$.
We assume a kinetic mixing between the hypercharge field $B_\mu$ and the $U(1)_X$ 
gauge field $C_\mu$ characterized by the mixing parameter $\delta$.
The mass growth for the gauge field occurs via the Stueckelberg mechanism and is  characterized
by the parameters $M_1$ and $M_2$. The scalar sector of the model contains the SM Higgs boson, $S$ and $\phi$. Therefore, the scalar potential is given by
\begin{align}
V&=V_{\rm SM}+\frac{1}{2}m^2_S S^2+\frac{\kappa_S}{3} S^3+\frac{\kappa_{\phi S}}{2}\phi^2 S+\frac{\lambda_S}{4} S^4+\frac{\lambda_\phi}{4} \phi^4+\frac{\lambda_{\phi S}}{2} \phi^2 S^2.
\label{potential}
\end{align} 
The mixing between the gauge field $B_\mu$ and $C_\mu$ in the Stueckelberg mass
mixing term will give  the dark fermions 
a millicharge, which is constrained by several experiments. For simplicity here we set $M_2=0$.
 After electroweak symmetry breaking and in the canonical basis where the 
  kinetic energies of the gauge fields are diagonalized and normalized,  
we have the following  set of couplings among particles of the  dark sector  and among particles of the dark sector and of the visible sector so that
\begin{align}
\label{D-darkphoton}
\Delta \mathcal{L}^{\rm int}&=\bar D\gamma^\mu(g^D_{\gamma'} A_{\mu}^{\gamma'}+g^D_{Z} Z_{\mu}+g^D_{\gamma} A_{\mu}^{\gamma})D
   + \frac{g_2}{2\cos\theta}\bar\psi_f\gamma^{\mu}\Big[(v'_f-\gamma_5 a'_f)A^{\gamma'}_{\mu}\Big]\psi_f,  
\end{align}
where $f$ runs over all SM fermions. In the limit of small kinetic mixing, $g_{\gamma'}^D\simeq  g_X Q_X$ which is a normal strength coupling. The couplings $g_{Z}^D$ and $g_{\gamma}^D$ are given in Appendix~\ref{app:B}. The vector and axial couplings are given by
\begin{equation}
\begin{aligned}
v'_f&=-\cos\psi[(\tan\psi-s_\delta\sin\theta)T_{3f}-2\sin^2\theta(-s_{\delta} \csc\theta+\tan\psi)Q_f],\\
a'_f&=-\cos\psi(\tan\psi-s_{\delta} \sin\theta)T_{3f},
\end{aligned}
\label{eqn:v-a}
\end{equation}
where $s_\delta= \sinh (\delta)$,
 $T_{3f}$ is the third component of isospin, $Q_f$ is the electric charge for the fermion and the angles $\theta$ and $\psi$ are defined in Appendix~\ref{app:B}.
 We also record here the couplings of the $Z_\mu$ and $A^\gamma_\mu$ in the canonically diagonalized basis which are given by~{\cite{Cheung:2007ut,Feldman:2007wj}
\begin{equation}
\Delta \mathcal{L}'_{\rm SM}=\frac{g_2}{2\cos\theta}\bar\psi_f\gamma^{\mu}\Big[(v_f-\gamma_5 a_f)Z_{\mu}\Big]\psi_f+e\bar\psi_f\gamma^{\mu}Q_f A^\gamma_{\mu}\psi_f\,.
\label{SMLag}
\end{equation}
Modifications to the visible sector interactions appear in the vector and axial couplings so that 
\begin{equation}
\begin{aligned}
v_f&=\cos\psi[(1+ s_\delta \tan\psi\sin\theta)T_{3f}-2\sin^2\theta(1+ s_\delta \csc\theta\tan\psi)Q_f],\\
a_f&=\cos\psi(1 + s_\delta \tan\psi\sin\theta)T_{3f}. 
\end{aligned}
\label{eqn:v-a}
\end{equation}
Further details can be found in refs.~\cite{Aboubrahim:2020lnr,Aboubrahim:2020afx,Aboubrahim:2021ycj,Aboubrahim:2021ohe}.
The free parameters of the model described above are subject to a number of  theoretical and experimental constraints. The theoretical constraints require that the model produce the desired
amount of dark matter, which in turn necessitates a careful analysis of chemical and kinetic decoupling
of different particle processes. It also requires that the model maintains the success of the
BBN analysis within the SM paradigm, i.e., that the model produces only negligible 
 extra relativistic degrees of freedom at BBN time, but makes a significant contribution 
  $\Delta N^{\rm}_{\rm eff}$ at the time the CMB was produced. These put severe constraints on the
  dark photon mass, the mass of the dark fermion, and on chemical and kinetic decoupling of 
  species in the dark sector. Additionally, the kinetic mixing and the dark photon mass are
  severely constrained by current experiments. Despite all the stringent constraints, we show that a considerable parameter space remains viable to construct a cosmologically consistent particle physics model that can help reduce
  the Hubble tension.

\section{Evolution  of hidden and visible sector temperatures} 
\label{sec:temp}

Hidden and visible sectors in general reside in heat baths
at different temperatures~\cite{Aboubrahim:2020lnr,Aboubrahim:2021ycj,Aboubrahim:2021dei,Foot:2014uba,Foot:2016wvj}. In  models of the type we consider, the bath for the  hidden sector is at
a much lower temperature than the bath for the visible sector. However,  because of the
coupling between the visible and the hidden sectors, their  temperature evolution is correlated.  Specifically, the entropy in each sector is not separately
conserved, but it is only the total entropy that is conserved.
Thus the total entropy is $S= \mathbb{s}R^3$, where $R$ is the scale factor and $\mathbb{s}=\mathbb{s}_v+\mathbb{s}_h$, where $\mathbb{s}_h$ is the hidden sector entropy density, $\mathbb{s}_v$ is the entropy density of the visible 
sector, and 
\begin{align}
\mathbb{s}&=\frac{2\pi^2}{45}\left(h_{\rm eff}^h T_h^3+h_{\rm eff}^v T^3\right).
\label{eq3.a}
\end{align}
Here $h^v_{\rm eff}$ and $h^h_{\rm eff}$ are the visible and the hidden effective entropy degrees of freedom and $T~(T_h)$ is the temperature of the visible (hidden) sector.
The conservation of the total 
entropy  leads to the evolution equation for the entropy density
\begin{align}
d\mathbb{s}/dt + 3 H\mathbb{s}=0.
\end{align}
The temperature evolution of the two sectors is also affected by the dependence
of the Hubble parameter on the temperatures of the two sectors. Thus 
 the Friedman equation gives
\begin{align}
H^2= \frac{8\pi G_N}{3} (\rho_v(T)  +\rho_h(T_h)),
\label{eq3.b}
\end{align}
where
 $\rho_v(T)$ and $\rho_h(T_h)$ are the energy densities of the visible and of the
hidden  sectors  and are given in terms of their respective energy density degrees of freedom as
\begin{align}
\rho_v&=\frac{\pi^2}{30}g_{\rm eff}^v T^4, ~~
\rho_h=\frac{\pi^2}{30}g_{\rm eff}^h T_h^4.
\label{rho-1}
\end{align}
For the visible sector effective degrees of freedom, $g^v_{\rm eff}$ and $h^v_{\rm eff}$, the recent tabulated results from lattice QCD~\cite{Drees:2015exa} are used, while for the hidden sector, $g^h_{\rm eff}$ and $h^h_{\rm eff}$ are given in the form of integrals in Appendix~\ref{app:A}.  The  time evolution of the energy density in the hidden sector is  given by             
\begin{align}
\frac{d\rho_h}{dt} +3H (\rho_h +p_h) =j_h,
\end{align}
where $p_h$ is hidden sector pressure density and $j_h$ is the source in the hidden sector. It encodes the heat exchange between the visible and the hidden sectors 
and is given by Eq.~(\ref{jh}) of Appendix~\ref{app:C}.

Since the temperatures of the visible and the hidden sectors are correlated,  a quantitative 
 analysis of this correlation is essential. For this reason we introduce a function $\eta=T/T_h$.
 The evolution equation for $\eta$ depends on the particle number densities, and so we need to identify the interactions
 that enter in the evolution. There are two such types of interactions.
 The first one involves
 the interactions of the dark particles among themselves, i.e., among $D, \bar D, \gamma', \phi$ and $S$.
  Then we have the feeble interactions between the  
  visible sector and the hidden sector, which involve processes of the type $i\bar i\to D\bar D$, and $i\bar i\to \gamma'$, where $i$ refers to the Standard Model particles 
  and the interactions are those induced by the kinetic energy mixing. We assume no initial number density for the hidden sector particles and it is only
  through the feeble interactions of the visible sector with the hidden sector that the hidden 
  sector particles are produced. Here we note that only $D, \bar D$ and $\gamma'$ 
  are produced by the portal interactions with the visible sector, while $\phi$ and $S$ are only produced by interactions within the dark sector.  The set of all interactions considered appear in the system of coupled Boltzmann equations, Eqs.~(\ref{yphi})$-$(\ref{yD}) (see Appendix~\ref{app:C}). Here we have introduced the dimensionless quantity $Y_a=n_a/\mathbb{s}$, the comoving number density or yield of a particle species $a$, and have written the equations relative to the hidden sector temperature $T_h$ instead of time. 

The ratio $\eta\equiv T/T_h$ is itself dependent on $T_h$, since we have taken $T_h$ as our reference temperature. Its evolution is driven by the energy injection from the SM into the dark sector. The energy exchange is given by the source term $j_h$.  In this 
   case one may derive the equation for $\eta$ so that 
\begin{align}
\frac{d\eta}{dT_h}= &-\frac{\eta}{T_h} +  \left[\frac{\zeta \rho_v+ \rho_h( \zeta-\zeta_h)+ j_h/(4H)}{\zeta_h\rho_h- j_h/(4H)}\right] \frac{d\rho_h/dT_h}{T_h (d\rho_v/dT)},
\label{eta}
\end{align} 
where $j_h$  is given by 
\begin{align}
\label{jh}
j_h=&\sum_i \Big[2Y^{\rm eq}_i(T)^2 J(i~\bar{i}\to D\bar{D})(T)+Y^{\rm eq}_i(T)^2 J(i~\bar{i}\to \gamma')(T)\Big]\mathbb{s}^2-Y_{\gamma'}J(\gamma'\to f\bar{f})(T_h)\mathbb{s}.
\end{align} 
The $J$ terms appearing in Eq.~(\ref{jh}) are given in Appendix~\ref{app:C}.	
In Eq.~(\ref{eta}), $\zeta$ and $\zeta_h$ are 1 for the radiation dominated era and 3/4 for
matter dominated era. More generally they are given by
 $\zeta= \frac{3}{4} (1+p_v/\rho_v)$ and $\zeta_h=\frac{3}{4} (1+p_h/\rho_h)$.

The relic density of $D$, which is dark matter,
 is related to  $Y^{\infty}_D$, the yield today, and the
entropy density today, $\mathbb{s}_0$ by
\begin{align}
\Omega h^2 = \frac{m_D Y^{\infty}_D \mathbb{s}_0 h^2}{\rho_c},
\label{relic}
\end{align}
where $\rho_c$ is the critical density, and $h=0.674$\cite{Planck:2018vyg}.
The  total relic density of dark matter as given by the
Planck Collaboration~\cite{Aghanim:2018eyx} is 
$(\Omega h^2)_{\rm PLANCK}=0.1198\pm 0.0012$. Note that since we are modifying the early cosmology of $\Lambda$CDM by adding extra radiation, the DM relic density is bound to change. This is taken into consideration in the numerical analysis as we explain later.

\section{Departure from kinetic equilibrium}\label{sec:keq}

As the temperature of the universe drops, annihilation processes (i.e., number-changing processes) become progressively inefficient. Once their rate falls below the Hubble expansion rate, the involved particle species undergo chemical decoupling. The onset of chemical decoupling between $D$ and $\gamma'$ in the dark sector initiates a dark freeze-out where the dark matter yield or comoving number density remains fixed with falling temperature. Despite this decoupling, particle species can remain in local thermal equilibrium through scattering processes with the thermal bath. Focusing on $D$, its temperature $T_D$ still follows that of the hidden sector's thermal bath $T_h=T_\phi$ as it scatters off relativistic species, namely $S$ and $\phi$, via the processes $DS\to DS$ and $D\phi\to D\phi$. The same holds for $S$: after its chemical decoupling, local thermal equilibrium can be maintained via $S\phi\to S\phi$. Even those processes will eventually lose their competition against the Hubble rate and decouple. Note that kinetic decoupling can occur before chemical decoupling in which case one needs to solve the full Boltzmann equations for the phase space distributions~\cite{Binder:2017rgn,Binder:2021bmg}. This has an effect on the relic density and is often numerically challenging. Here we restrict ourselves to the case where kinetic decoupling happens much after chemical decoupling has occurred which is the assumption made when writing the Boltzmann equations, Eqs.~(\ref{yphi})$-$(\ref{yD}).      

To determine the temperature of kinetic decoupling of a dark sector species $X$ with mass $m_X$, one needs to solve the second moment of the Boltzmann equation. An analytical solution is quite complicated but can be simplified by assuming that the scattering particle is non-relativistic which means that $T_X\ll m_X$ and that the typical momentum transfer in the scattering processes is small compared to $m_X$. Keeping only leading order terms in $p^2/m_X^2$, we write the Boltzmann equation as~\cite{Bringmann:2009vf,Bringmann:2006mu}
\begin{equation}
(\partial_t+5H)T_X=2m_X\gamma(T_\phi)(T_\phi-T_X),
\label{2ndmoment}
\end{equation}
where $\gamma(T_\phi)$ is the momentum transfer rate evaluated at the bath temperature $T_\phi$, $T_X$ is the temperature of particle $X$ (used to describe departure from kinetic equilibrium) with number density $n_X$,  and is defined by
\begin{equation}
T_X=\frac{1}{3m_X n_X}\int\frac{d^3p}{(2\pi)^3}p^2 f(p),
\end{equation}
where $f(p)$ is
the one particle phase space distribution and the Hubble parameter is given by 
\begin{equation}
H^2(T_\phi)=\frac{4\pi^3}{45 M^2_{\rm Pl}}(g_{\rm eff}^v\eta^4+g^h_{\rm eff})T^4_\phi.
\end{equation}
The momentum transfer rate in Eq.~(\ref{2ndmoment}) is given in terms of an average of the matrix element squared over the Mandelstam variable $t$~\cite{Gondolo:2012vh}
\begin{equation}
\gamma(T_\phi)=\frac{g_\phi}{384\pi^3 m_X^4 T_\phi}\int dE \,f^{\pm}(E)(1\mp f(E))\int_{-4k^2}^0 (-t)|\mathcal{M}|^2\,dt,
\end{equation}
where $E$ is the energy of the bath particle $\phi$ with $g_\phi=1$. The upper sign in $f^{\pm}(E)(1\mp f(E))$ correspond to fermions and the lower sign to bosons. The temperature of kinetic decoupling $T_{\rm kd}$ is calculated as
\begin{equation}
\gamma(T_\phi)=H(T_\phi)\Big|_{T_\phi=T_{\rm kd}},
\label{kd}
\end{equation}
assuming instantaneous decoupling. Prior to decoupling, $T_X$ traces $T_\phi$ as both temperatures drop as $1/R$. However, after decoupling, the temperature of $X$ ($D$ and $S$ in our model) decreases as $1/R^2$, while $T_\phi$ continues as $1/R$ assuming that there are no further interactions or decays. This situation will change in the next section when we discuss out-of-equilibrium decay of $S$, which adds to the energy density of $\phi$. An illustration of how Eq.~(\ref{kd}) works 
is given in Fig.~\ref{fig1}, where we show the rate of the processes $D\phi\to D\phi$, $DS\to DS$ (blue curve) and $\phi S\to \phi S$ (gray curve) along with the Hubble parameter (dashed line) plotted against the visible sector temperature. Kinetic decoupling occurs when the rate falls below $H(T)$ which for $D$ occurs at a higher temperature than that for $S$ for this particular choice of parameters (benchmark (b) of Table~\ref{tab1} below).  

Note that our model predicts late kinetic decoupling, while there exists a lower limit on $T_{\rm kd}$ from Lyman-$\alpha$ forest measurements~\cite{Bringmann:2016ilk} implying that $T_{\rm kd} >100$ eV. This will set constraints on the couplings in the dark sector.

\begin{figure}[t]
    \centering
  \includegraphics[width=0.6\textwidth]{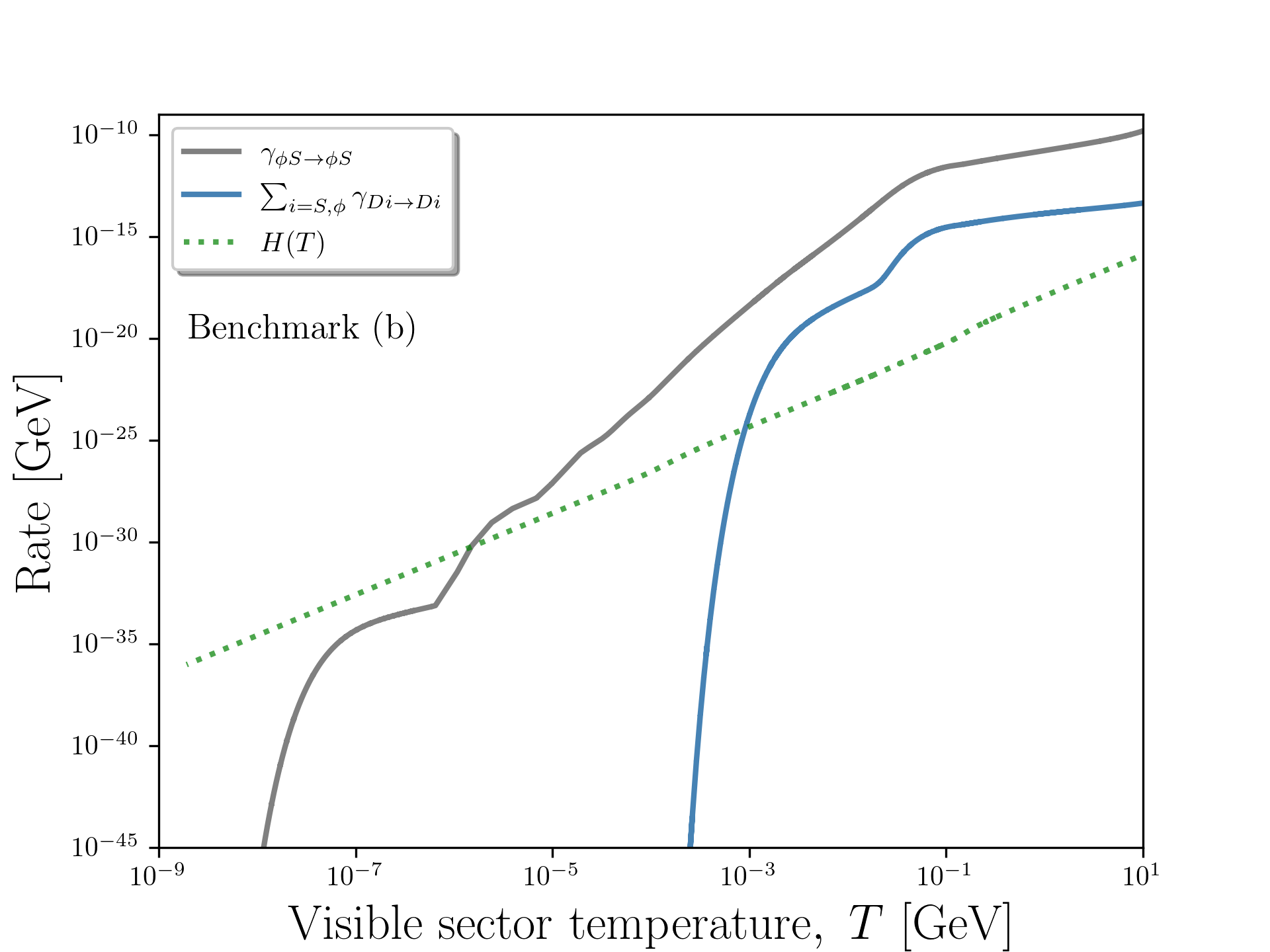}
    \caption{Momentum transfer rates for $D\phi\to D\phi$, $DS\to DS$ (blue curve) and $\phi S\to \phi S$ (gray curve) along with the Hubble rate (dashed line) as a function of the visible sector temperature. For this set of parameters corresponding to benchmark (b) in Table~\ref{tab1}, $T^D_{\rm kd}>T^S_{\rm kd}$.}
    \label{fig1}
\end{figure}

\section{Numerical analysis: yield and temperature evolution prior to decoupling}\label{sec:numerical}

In this section we discuss the results pertaining to temperature and number density evolution before the full decoupling of the dark sector species. Here we numerically solve the coupled Boltzmann equations, Eqs.~(\ref{yphi})$-$(\ref{yD}), together with the temperature evolution of the two sectors, Eq.~(\ref{eta}), in order to track the growth of the particle yields as a function of the temperature $T_h$. The system of equations form a set of stiff differential equations which are solved in two steps: (1) integrating from a high temperature assuming $T_h/T\ll 1$ until chemical decoupling sets in the dark sector at $T_{\rm cd}$ and (2) integrating $dY_S/dT_h$ and $dY_\phi/dT_h$ from $T_{\rm cd}$ down to 1 eV assuming $Y_D=\text{constant}$ and $Y_{\gamma'}=0$. In every calculation we verified that $Y_{\gamma'}$ is small enough to be discarded (since $\gamma'$ decays away before BBN) and that $D$ has already undergone a dark freeze-out. Splitting the temperature range this way helps the ordinary differential equation solver tackle the stiffness of the system at hand. 

\begin{table}[t]
\caption{\label{tab1}
Selection of benchmarks used in this analysis. We have set $\epsilon=M_2/M_1=0$, $\kappa_S=10^{-18}$ GeV, $y_\phi=0$ and $g_X=0.0125$. All masses are in GeV.}
\begin{center}
\begin{tabular}{|ccccccccc|}
\hline\hline
Model & $m_D$ & $m_{\gamma'}$ & $ m_S$ & $\delta$  & $y_S$ & $\kappa_{\phi S}$ [GeV] & $\lambda_{\phi S}$ & $\Delta N_{\rm eff}^{\rm CMB}$ \\
\hline
(a) & 0.1 & 0.9 & $10^{-2}$ & $4.6\times 10^{-11}$ & $3.0\times 10^{-3}$ & $1.9\times 10^{-18}$ & $1.0\times 10^{-7}$ & 0.43 \\
(b) & 0.3 & 0.2 & $10^{-2}$ & $1.6\times 10^{-9}$ & $1.0\times 10^{-3}$ & $6.0\times 10^{-18}$ & $5.0\times 10^{-6}$ & 0.55 \\
(c) & 0.6 & 0.5 & $10^{-3}$ & $6.0\times 10^{-10}$ & $3.0\times 10^{-3}$ & $3.3\times 10^{-19}$ & $5.0\times 10^{-7}$ & 0.36 \\
(d) & 1.0 & 0.3 & $10^{-2}$ & $1.7\times 10^{-9}$ & $5.0\times 10^{-3}$ & $8.5\times 10^{-19}$ & $1.0\times 10^{-8}$ & 0.54 \\
\hline\hline
\end{tabular}
\end{center}
\label{tab1}
\end{table}

We select a set of representative benchmarks which satisfy all the experimental constraints on the dark matter relic density and the dark photon mass and which can help alleviate
the Hubble tension. They are given in Table~\ref{tab1}. Note that a more general investigation of the available parameter space will be given later in this section and in the next sections as well. The DM relic density for benchmarks (b)$-$(d) is $\sim 0.12$ while that for (a) is $0.07$ which can be explained by the smallness of the kinetic mixing $\delta$. We note here that $\kappa_s$ and  $\kappa_{\phi s}$
have dimension of mass and one expects these to be Planck suppressed so that the first such term is of
size $\lambda m^2_{\rm EW}/m_{\rm Pl}$. For $m_{\rm EW}\sim 10^{2}$ GeV, $m_{\rm Pl}\sim 10^{19}$ GeV, and $\lambda\sim 10^{-3}$ one gets $\kappa\sim \kappa_{\phi s}\sim 10^{-18}$ GeV.

\begin{figure}[H]
    \centering
 \includegraphics[width=0.495\textwidth]{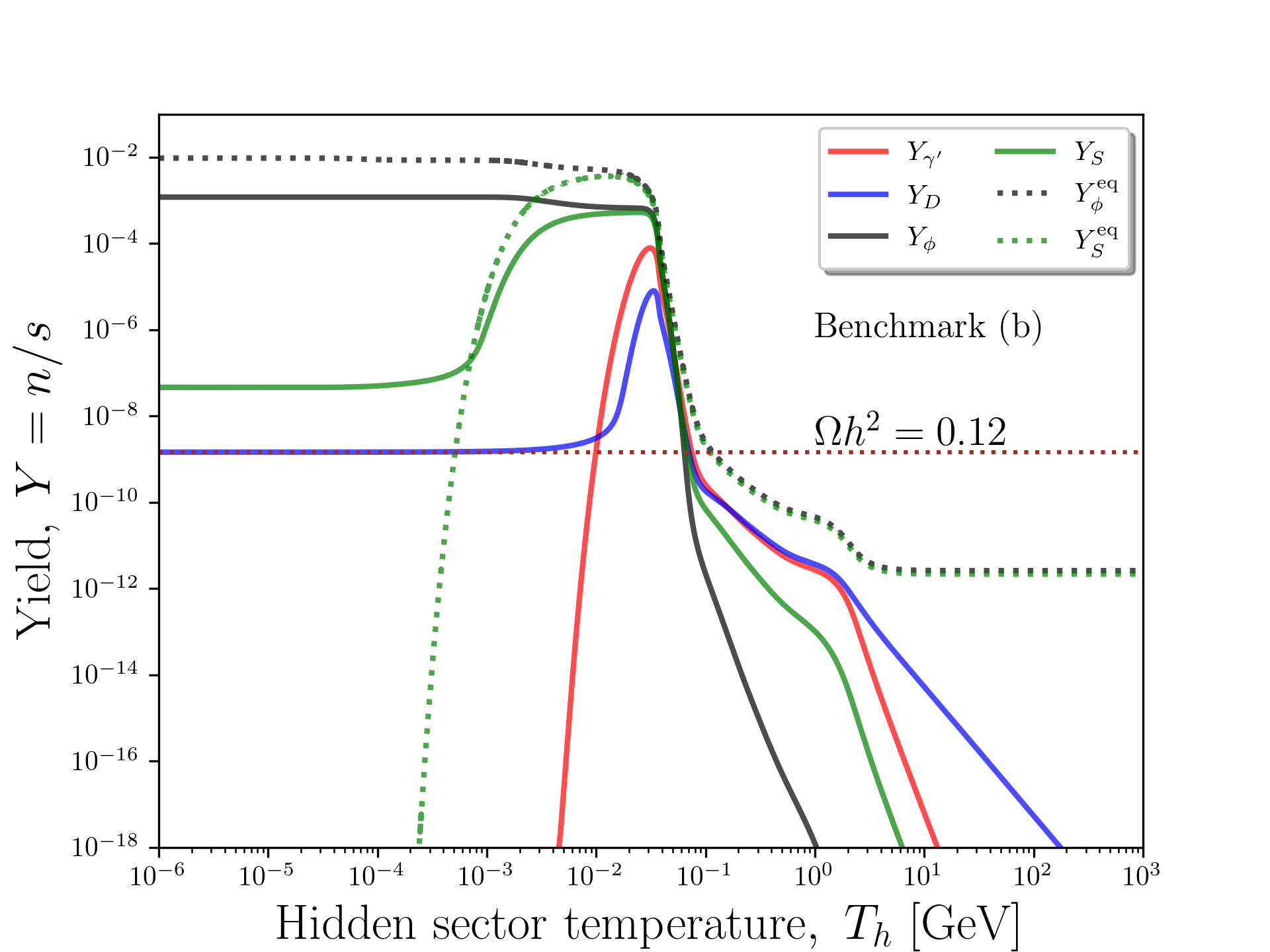}
 \includegraphics[width=0.495\textwidth]{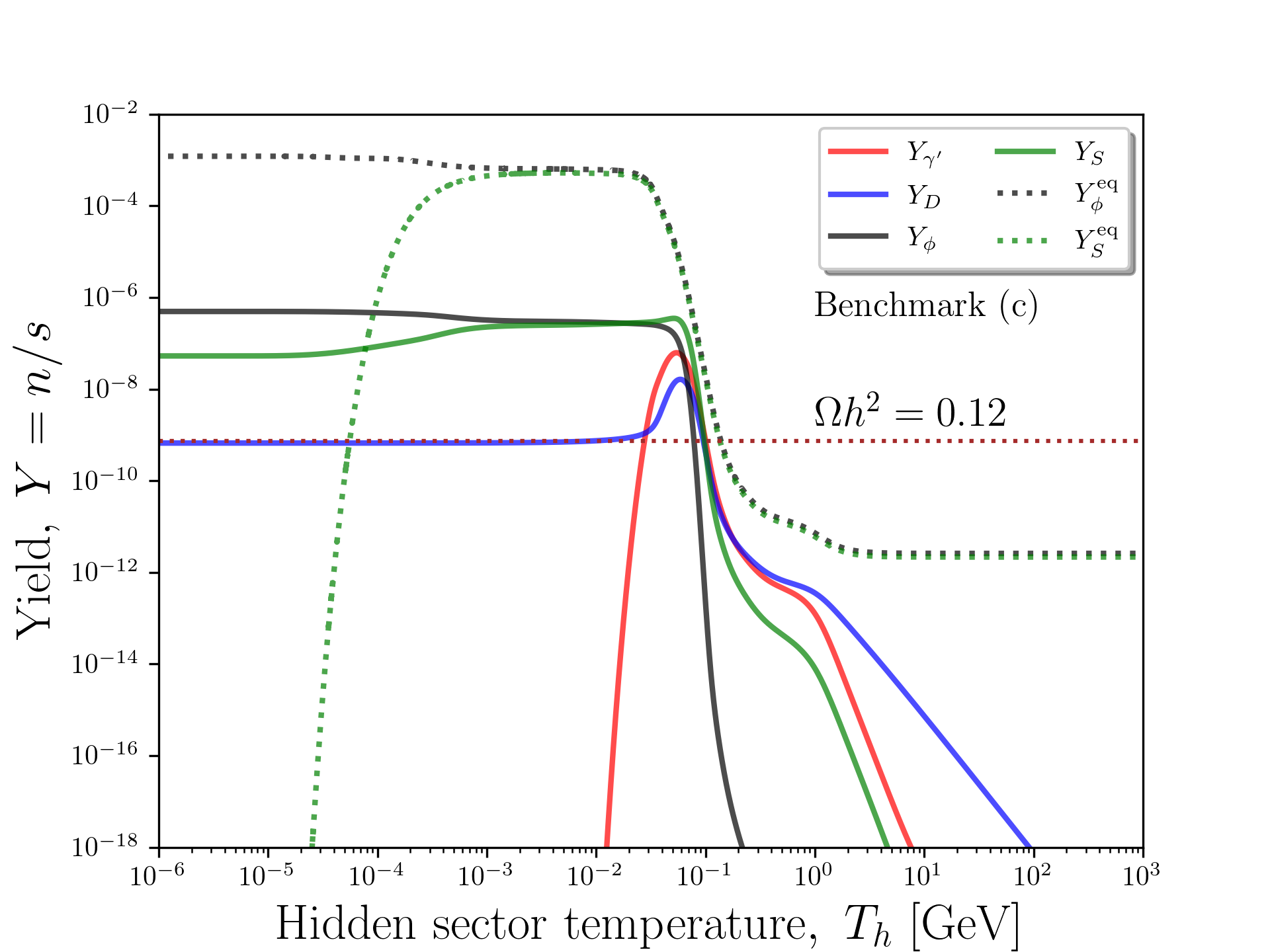}
    \caption{Evolution of the hidden sector particle yields (solid lines) as a function of the hidden sector temperature for benchmark (b) (left panel) and benchmark (c) (right panel) of Table~\ref{tab1}. The dashed curves correspond to the equilibrium distribution of the particles $S$ and $\phi$. Here $(\Omega h^2)_D\simeq 0.12$. }
    \label{fig2}
\end{figure}

We show in Fig.~\ref{fig2} the evolution of the particle yields as a function of the hidden sector temperature for benchmarks (b) (left panel) and (c) (right panel). The solid lines correspond to the particle yields as obtained from solving the Boltzmann equations, while the dotted lines correspond to the equilibrium yields of $\phi$ and $S$. One can see from the left panel that the particle yields increase gradually with $D$ and $\gamma'$ taking the lead as they are produced by SM processes while the yields for $S$ and $\phi$ are smaller since they are produced only via their interactions with $D$ and with each other (see Eqs.~(\ref{yphi}) and~(\ref{yS})). The yields undergo a sudden surge at $T_h\sim 0.1$ GeV which happens near the QCD crossover ($T=\eta T_h < 1$ GeV) where the freeze-out of the SM degrees of freedom causes an entropy dump leading to a sudden drop in the visible sector temperature relative to the hidden sector. One can see in the left panel of Fig.~\ref{fig3} that the visible temperature drops sharply at $T\sim 1$ GeV while the hidden sector temperature drops at a slower rate. This can be seen using the conservation of entropy
\begin{align}
S&=\frac{2\pi^2}{45}[h^v_{\rm eff}\eta^3+h^h_{\rm eff}]T^3_h R^3=\frac{2\pi^2}{45}[h^v_{\rm eff}+h^h_{\rm eff}\xi^3]T^3 R^3=\text{const.}
\label{stv}
\end{align}
Not only does $h^v_{\rm eff}$ undergo a sudden drop, but so does $\eta$ (see right panel of Fig.~\ref{fig4}). As a result, one can immediately see from the first equality of Eq.~(\ref{stv}) that $T_h$ drops at a slower rate. On the contrary, the second equality of Eq.~(\ref{stv}) shows that a drop in $h^v_{\rm eff}$ is partially compensated by an increase in the second term proportional to $\xi$, where we define $\xi\equiv 1/\eta$ (see left panel of Fig.~\ref{fig4}). As a result, the drop in the visible sector temperature is not slowed down.  Following this increase, the yield of $\gamma'$ starts decreasing (red curve) as the decay rate overtakes the Hubble parameter (see black curve in the right panel of Fig.~\ref{fig3}). At the same time dark matter annihilation in the dark sector becomes very efficient, leading to a drop in the dark matter yield (blue curve) before freeze-out sets in as the rate of the process $D\bar{D}\to\gamma'\gamma'$ drops below the Hubble rate. At this point, $D$ and $\gamma'$ are chemically decoupled after they have been in chemical equilibrium for some time, as seen in the right panel of Fig.~\ref{fig3}, leading to dark matter freeze out and the yield leveling off. Note that this happens at $T_h\sim 10^{-2}$ GeV which is in the neighborhood of $m_D/20$, the standard freeze-out temperature.

\begin{figure}[t]
    \centering
  \includegraphics[width=0.495\textwidth]{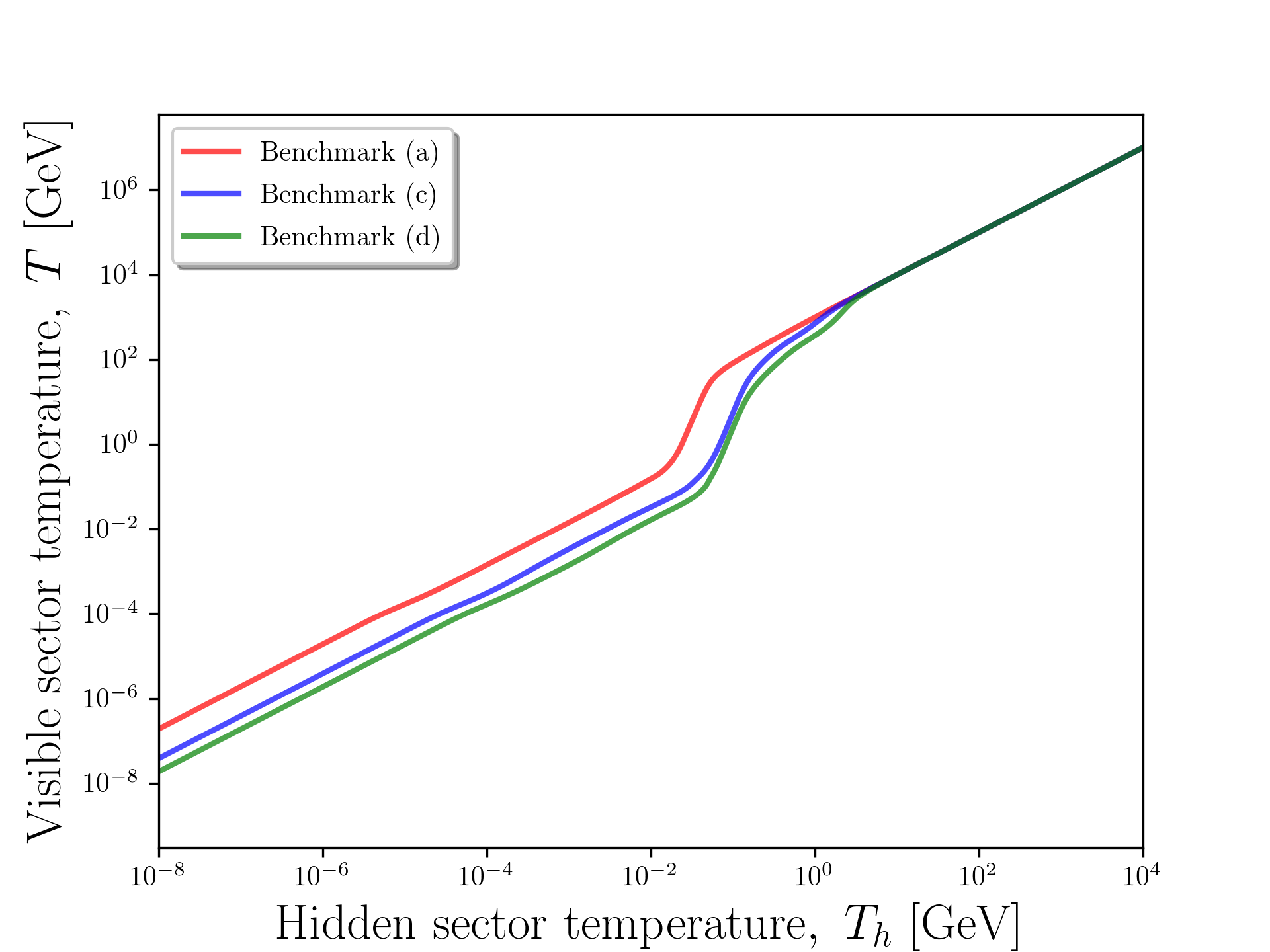}
   \includegraphics[width=0.495\textwidth]{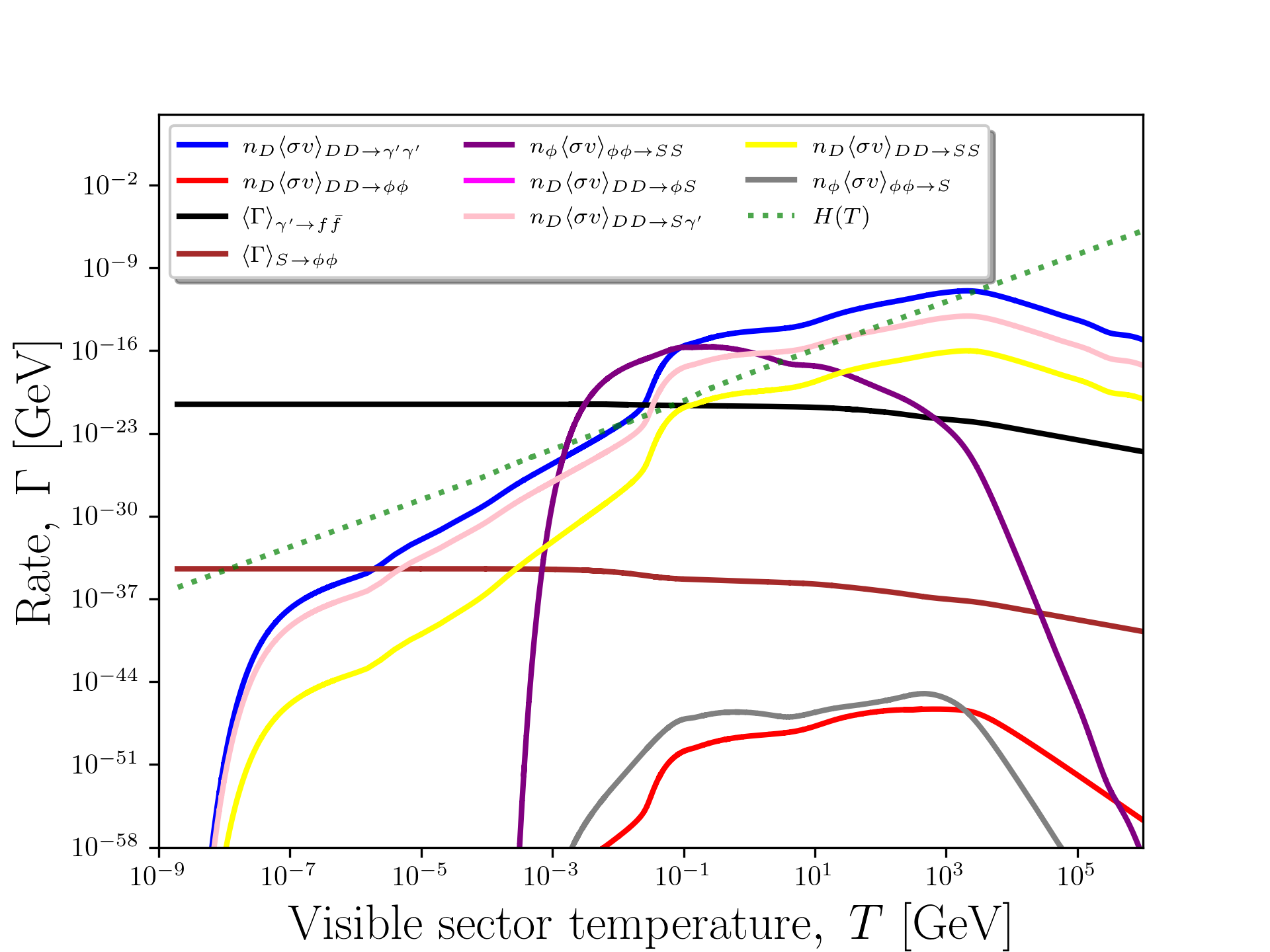}
    \caption{Left panel: Variation of the visible sector temperature as a function of the hidden sector temperature for benchmarks (a), (c) and (d) of Table~\ref{tab1}. Right panel: Rates for the different processes that enter into the Boltzmann equation (solid curves) as compared to the Hubble rate (dashed line) plotted as a function of the visible sector temperature.}
    \label{fig3}
\end{figure}

We also notice from the right panel of Fig.~\ref{fig3} that $\phi$ and $S$ enter into chemical equilibrium (purple curve) as well as $D$, $\gamma'$ and $S$ (pink curve). 
Thus there is a period in the thermal history of the dark sector where all species are in thermal equilibrium. Chemical decoupling between $\phi$ and $S$ happens after chemical decoupling of $D$ and $\gamma'$ and so the freeze-out of $S$ happens at a lower temperature. From the left panel of Fig.~\ref{fig2}, the yield of $S$ can be seen leveling off at $T_h\sim 5\times 10^{-4}$ GeV. The drop in the yield of $S$ is followed by a rise in $\phi$ before the yield of the latter continues at a fixed value. Notice that neither $S$ nor $\phi$ follow their equilibrium distribution (plotted as dotted curves) which is an essential criterion in our analysis to avoid the BBN constraint on $\Delta N_{\rm eff}$ as we discuss later.  We note also that $S$ does not contribute to the relic density, since it will eventually decay near recombination as the rate of $S\to\phi\phi$ overtakes the Hubble rate (see right panel of Fig.~\ref{fig3}).

\begin{figure}[H]
    \centering
  \includegraphics[width=0.495\textwidth]{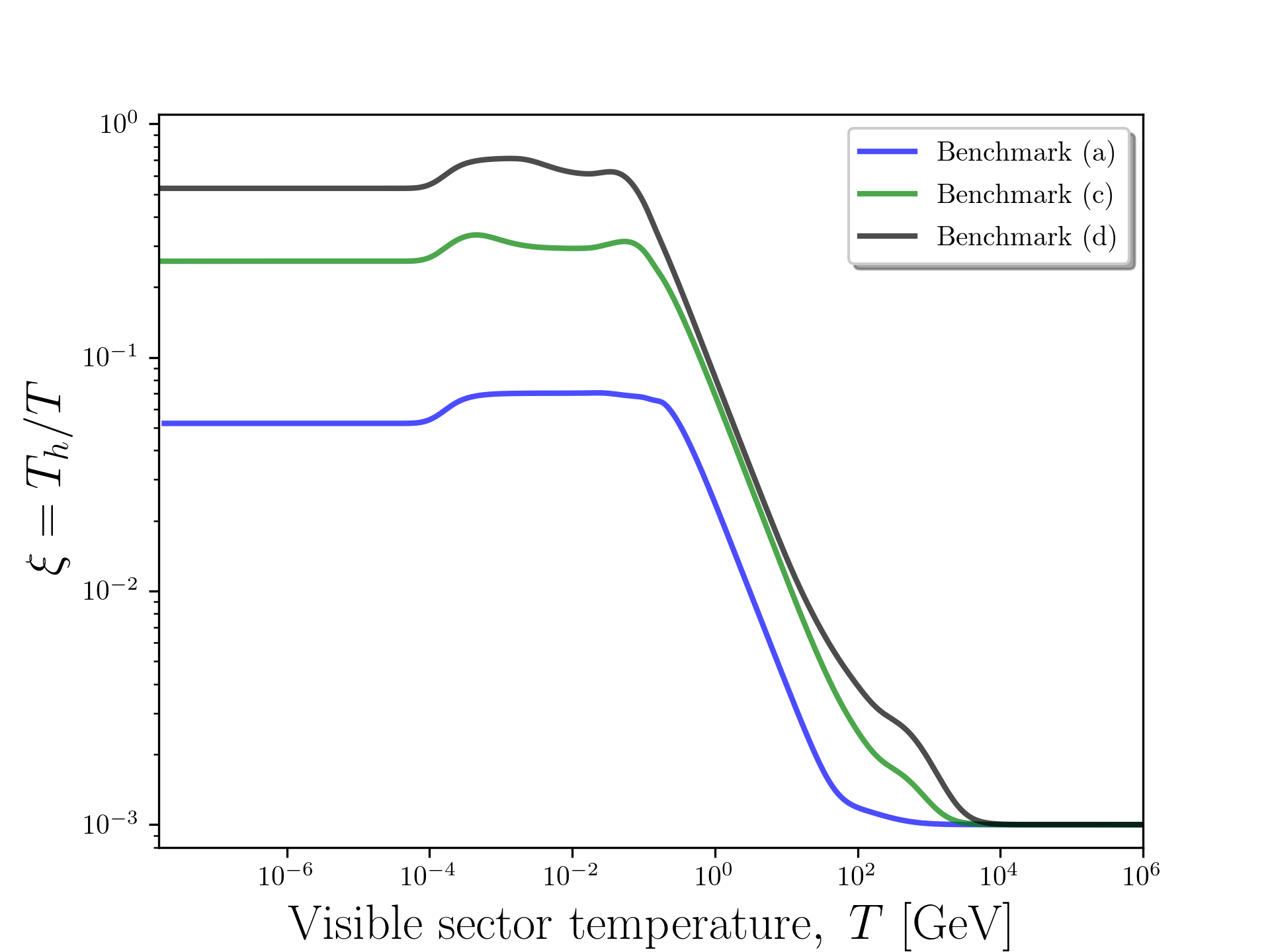}
   \includegraphics[width=0.495\textwidth]{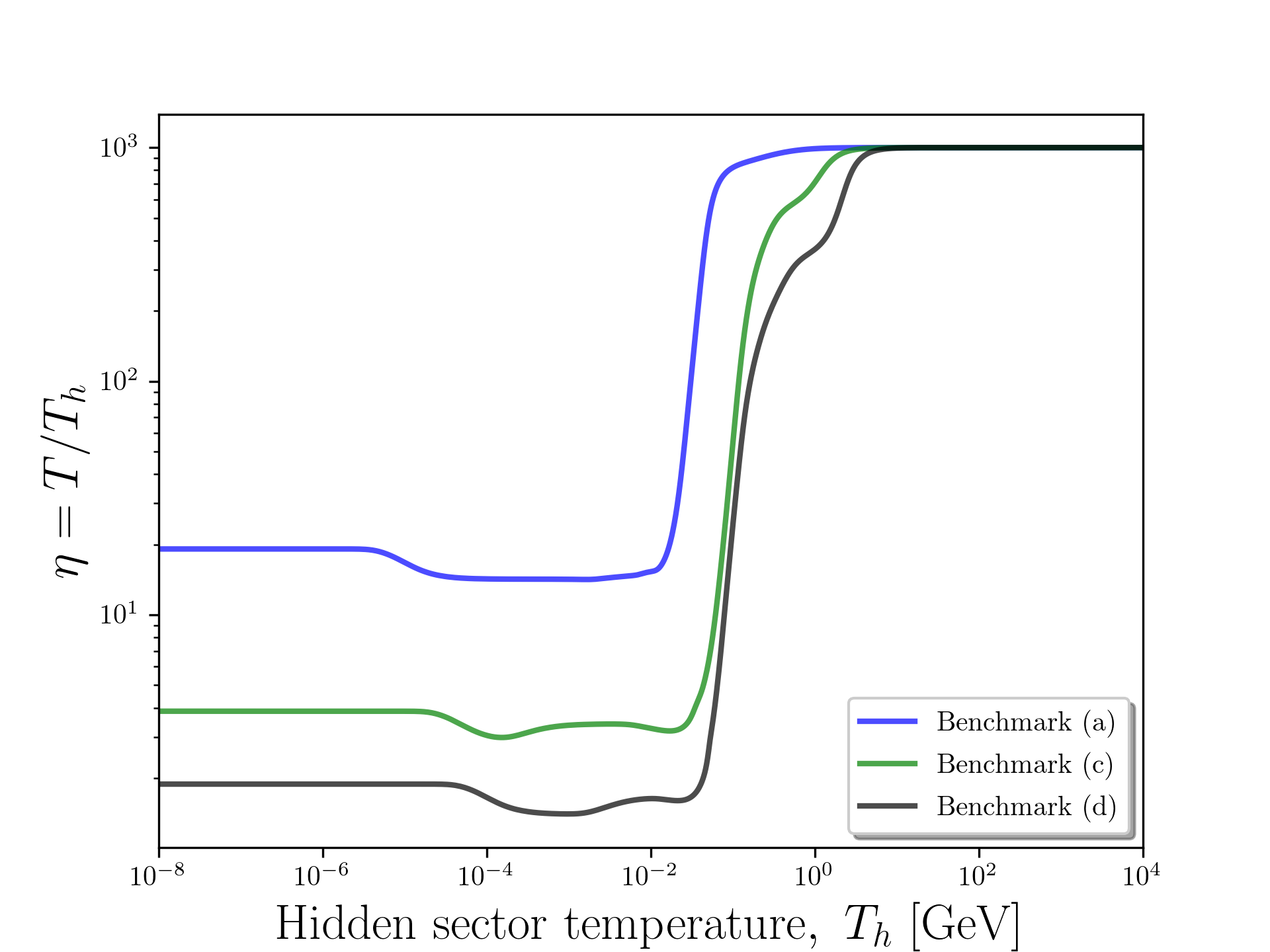}
    \caption{Left panel: Evolution of the hidden sector temperature represented by $\xi\equiv 1/\eta$ as a function of the visible sector temperature for benchmarks (a), (c) and (d) of Table~\ref{tab1}. Right panel: The ratio $\eta$ as a function of the hidden sector temperature for the same benchmarks. }
    \label{fig4}
\end{figure}

The energy injection from the SM into the cold dark sector gradually causes its temperature to increase. This is depicted in the left panel of Fig.~\ref{fig4} for benchmarks (a), (c) and (d) of Table~\ref{tab1}. For those benchmarks, the two sectors do not thermalize and $\xi$ reaches at most a value $\sim 0.55$ (for benchmark (d)) after BBN. The degree at which the two sectors reach thermal equilibrium depends on the size of the kinetic mixing $\delta$ which is why benchmark (a) has the smallest $\xi$ value since it has the smallest $\delta$.   A larger value of $\delta$ can produce a thermal distribution of $\phi$, and for $\xi$ large enough contributions to $\Delta N_{\rm eff}$ at BBN can be sizable which implies that $\delta$ is constrained in order to maintain the successful BBN predictions of the SM. 

We now discuss the available parameter space of the model taking into account the dark matter relic density constraint and limits on the dark photon mass. For a fixed value of $m_D$ and $g_X$, we scan the parameter space in the kinetic mixing-dark photon mass plane. We only keep points that satisfy the dark matter relic density, $(\Omega h^2)_D\sim$ 0.1$-$0.125\footnote{The presence of extra radiation causes the DM relic density to increase so that matter-radiation equality is maintained at a fixed redshift. The parameters of our model offer a flexibility when it comes to the dark matter relic density.}, and ones where kinetic decoupling of $D$ happens after its chemical decoupling. We also remove points with very late kinetic decoupling of $S$ and those with $\Delta N_{\rm eff}^{\rm BBN}>0.2$.  Further, there is an array of experimental limits on the dark photon mass which mixes kinetically with the SM sector. These include CHARM~\cite{Bergsma:1985qz,Tsai:2019mtm}, 
E141~\cite{Riordan:1987aw} and $\nu$-CAL~\cite{Blumlein:1990ay,Blumlein:1991xh,Tsai:2019mtm,Blumlein:2013cua,Blumlein:2011mv},
E137~\cite{Andreas:2012mt,Bjorken:2009mm}, electron and muon $g-2$~\cite{Endo:2012hp}, BaBar~\cite{Lees:2014xha}, NA48~\cite{Batley:2015lha}, NA64~\cite{Banerjee:2018vgk,Banerjee:2019hmi}, E141~\cite{Riordan:1987aw} and LHCb~\cite{LHCb:2017trq,LHCb:2019vmc}. The limits are obtained from \code{darkcast}~\cite{Ilten:2018crw}. Additionally, more stringent constraints arise
 from Supernova SN1987A~\cite{Chang:2016ntp} and BBN. The experimental limits as well as the region resulting from a scan of our model are presented in Fig.~\ref{fig5}.  In the top left panel of Fig.~\ref{fig5} we exhibit  the case for  $m_D=0.1$ GeV,  and in the top right
  panel the case for   $m_D=0.3$ GeV.  In  bottom left panel we exhibit  the case for  $m_D=0.6$ GeV,  and in the bottom right panel the case for   $m_D=1.0$ GeV  where for all cases we have  $g_X=0.0125$. The analysis of the panels shows that a part of the parameter space in the range $m_{\gamma'}=0.1$ GeV to 1.0 GeV is available. Further, in the top panels of Fig.~\ref{fig5}, one finds two allowed regions consistent with all constraints for $\delta\sim\mathcal{O}(10^{-7})$. These regions are in the vicinity of $m_{\gamma'}=2m_D$ which activates the channel $D\bar{D}\to\gamma'$ causing a drop in the dark matter relic density. To compensate this, an increase in $\delta$ is required.

\begin{figure}[H]
    \centering
    \includegraphics[width=0.495\textwidth]{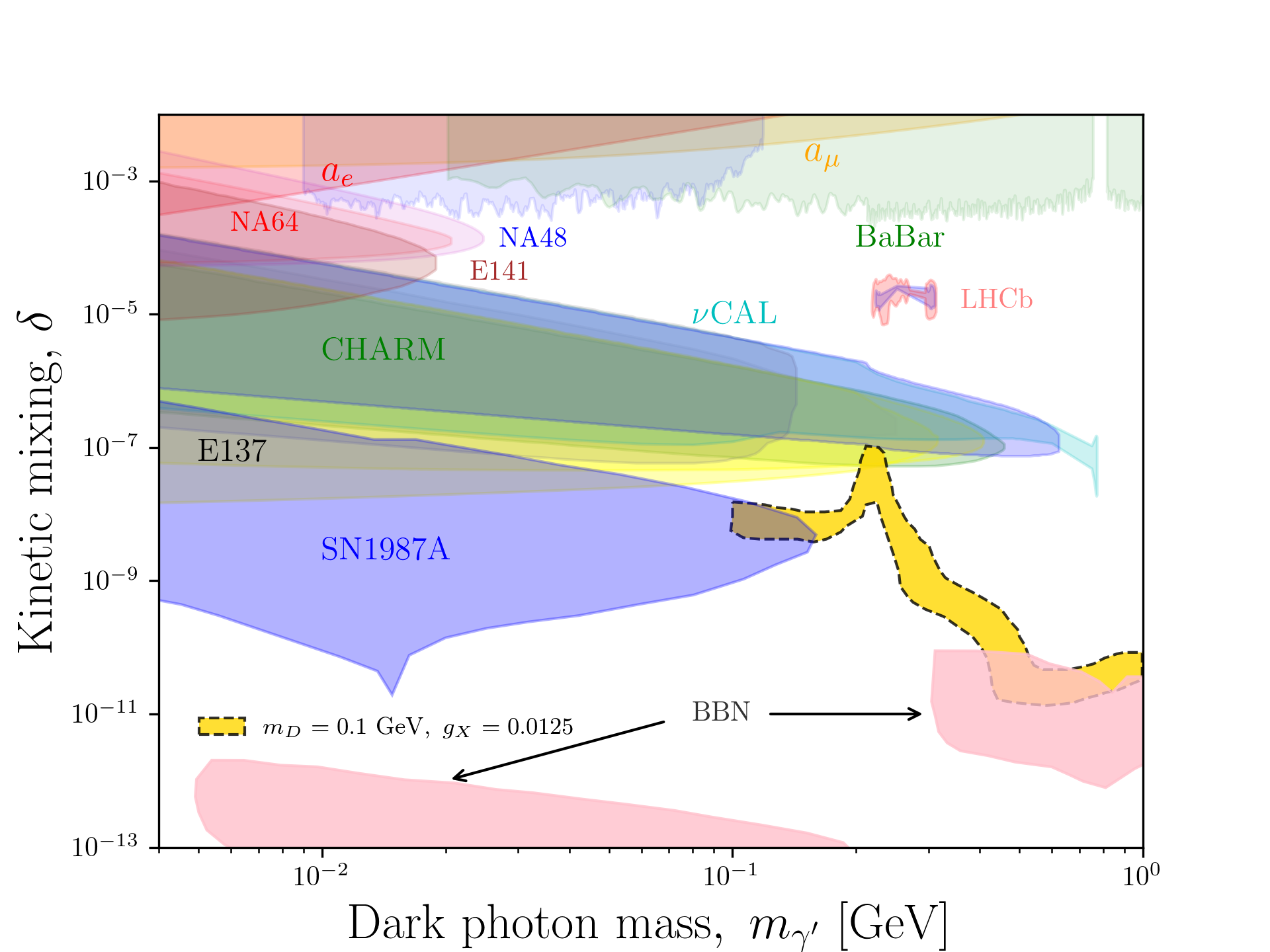}
  \includegraphics[width=0.495\textwidth]{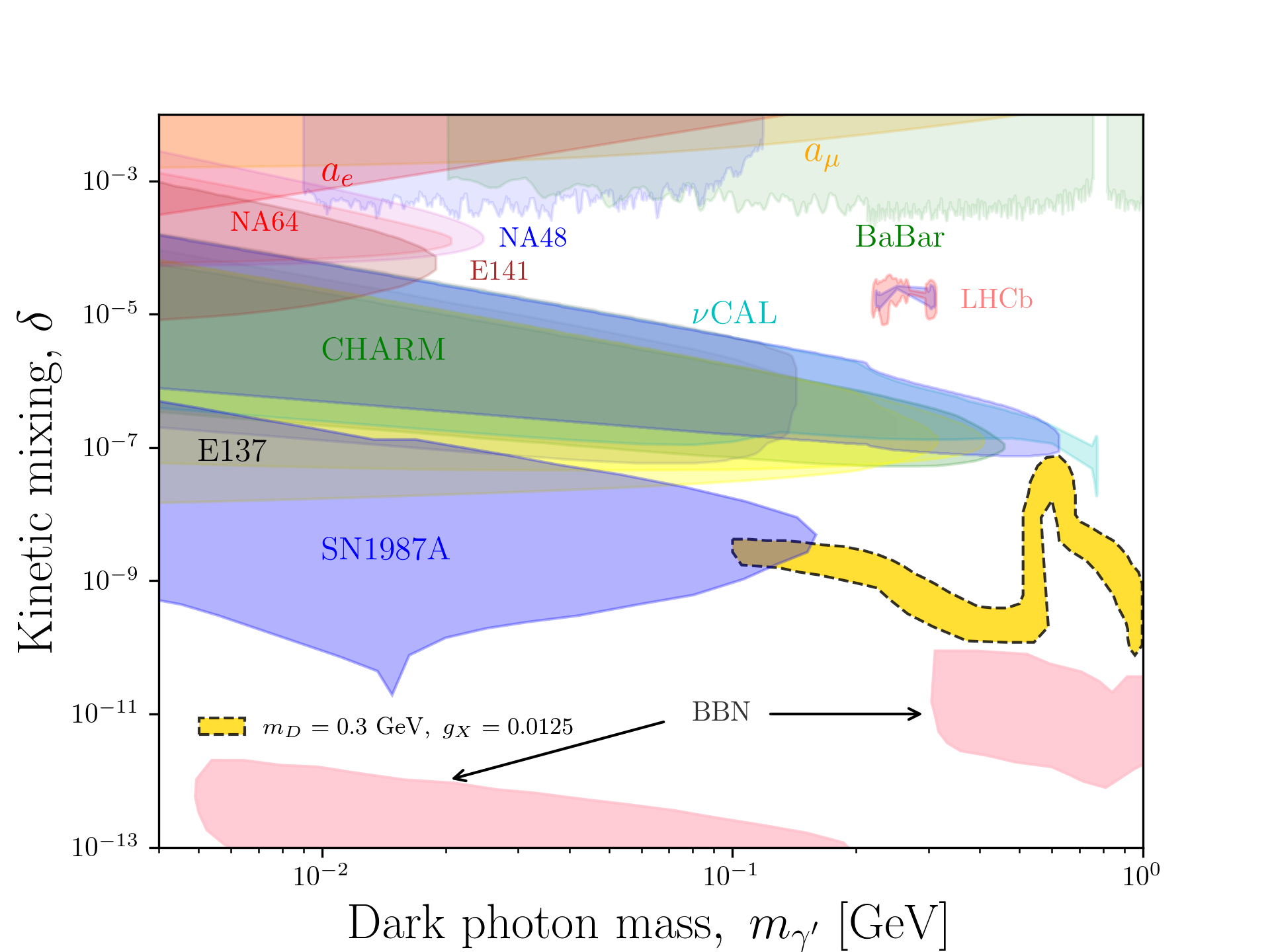}\\
    \includegraphics[width=0.495\textwidth]{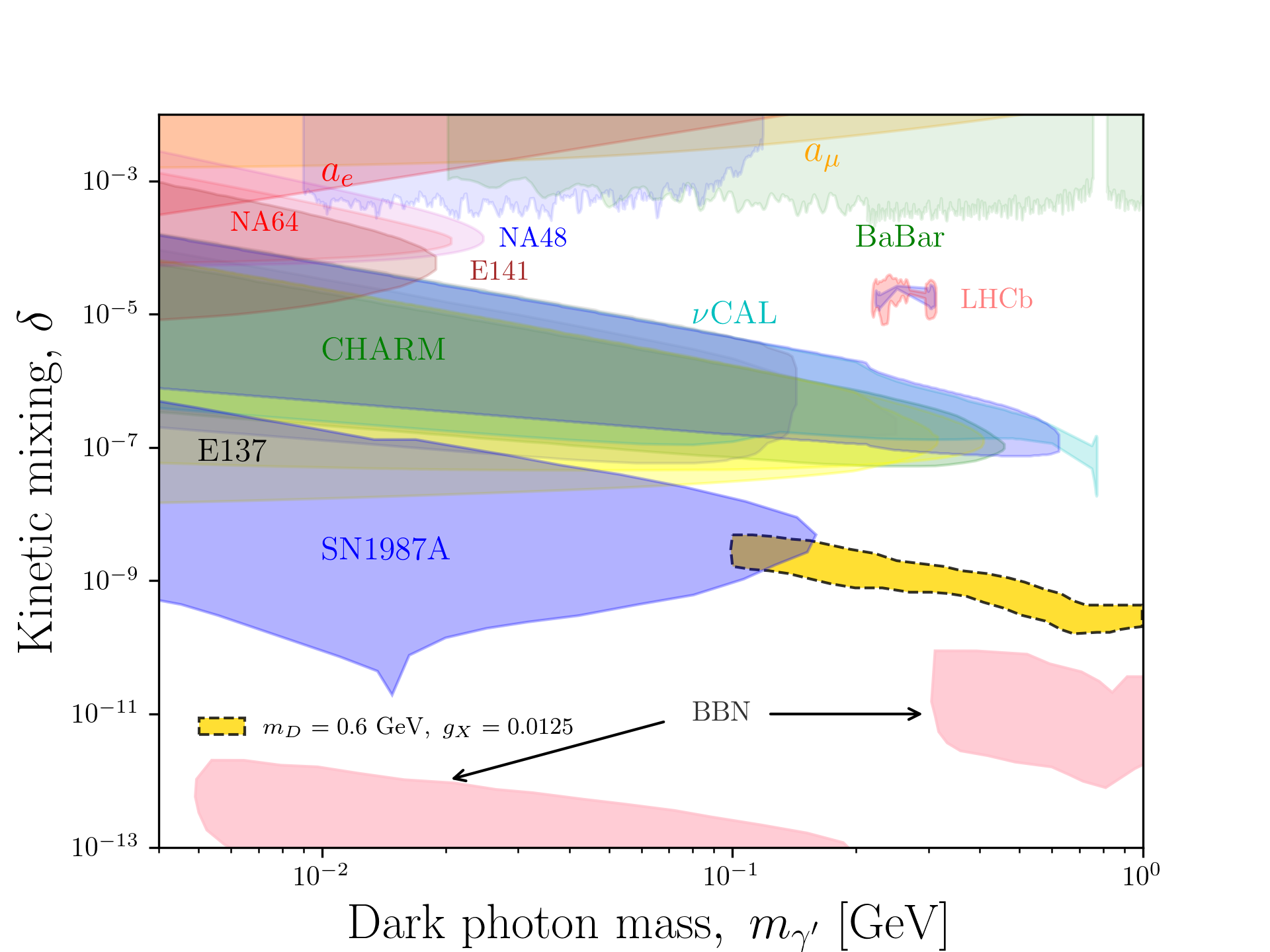}
  \includegraphics[width=0.495\textwidth]{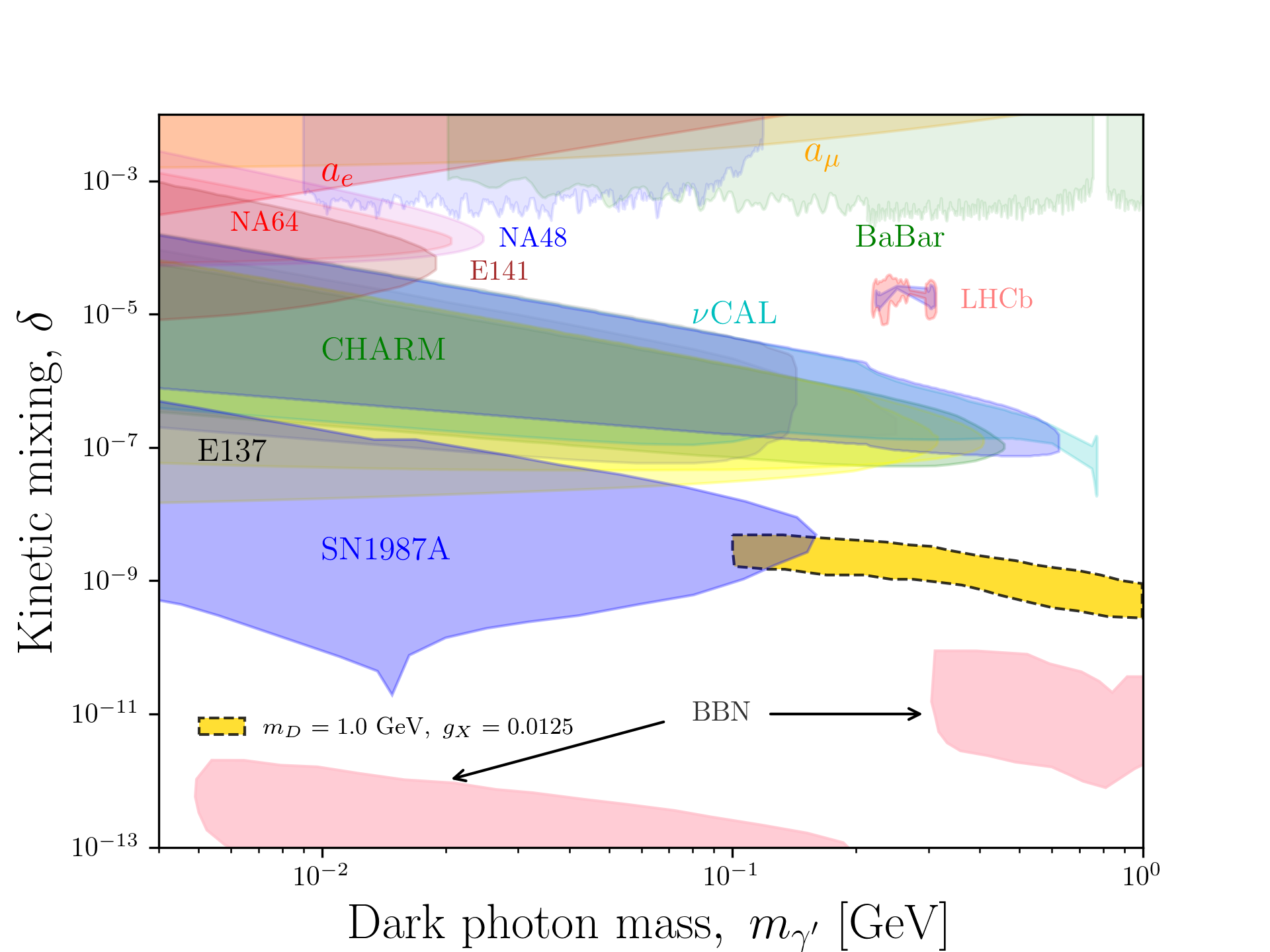}
    \caption{Exhibition of the $2\sigma$ regions in the space of the kinetic mixing parameter
    $\delta$ and the dark photon mass $m_{\gamma'}$ for four dark fermion masses: 
    $m_D=0.1$ GeV (top left panel), $m_D=0.3$ GeV (top right panel),  $m_D=0.6$ GeV (bottom left panel),  and $m_D=1.0$ GeV (bottom right panel). Other parameters are as shown in Table~\ref{tab1} for benchmark (a). The regions excluded by  several experiments as discussed in text are also
      exhibited. The analysis shows that there exists a significant region of the parameter space
      consistent with the experimental constraints
      which may help alleviate the Hubble tension.    }
    \label{fig5}
\end{figure}

\section{Contributions to $\Delta N_{\rm eff}$}\label{sec:neff}

After neutrino decoupling from photons, the visible sector has two separate thermal baths: the neutrinos with temperature $T_\nu$ and the photons with temperature $T_\gamma$, where $T_\gamma=(11/4)^{1/3}T_\nu$ based on entropy conservation. Note that in principle, the three neutrino species have different temperatures, however, neutrino oscillations tend to thermalize the three flavors and one can describe them by a single temperature. The effective number of relativistic degrees of freedom in the SM is thus given by
\begin{equation}
N_{\rm eff}^{\rm SM}=\frac{8}{7}\left(\frac{11}{4}\right)^{4/3}\left(\frac{\rho-\rho_\gamma}{\rho_\gamma}\right)=3\left(\frac{11}{4}\right)^{4/3}\left(\frac{T_\nu}{T_\gamma}\right)^4.
\end{equation}
Precision calculations of $N_{\rm eff}$ give a value slightly different than three, namely $N_{\rm eff}^{\rm SM}=3.046$~\cite{Mangano:2005cc,deSalas:2016ztq} after taking into account non-instantaneous decoupling of neutrinos, finite temperature corrections, neutrino oscillations and effects of $e^+e^-$ annihilation to neutrinos. Deviations from the SM value of $N_{\rm eff}$ can be realized in beyond the SM physics with particles contributing to the energy density of the universe. At BBN time, an increase in $N_{\rm eff}$ (and consequently in the energy density) would increase the expansion rate of the universe leading to observable effects on the light elements' abundances which are successfully predicted by BBN (in particular the Helium mass fraction). Thus, extra relativistic degrees of freedom in a particle physics model must not spoil BBN predictions. Deviations from $N_{\rm eff}^{\rm SM}$ are written as 
\begin{equation}
\Delta N_{\rm eff}\equiv N_{\rm eff}-N_{\rm eff}^{\rm SM}=\frac{8}{7}\left(\frac{11}{4}\right)^{4/3}\frac{\rho_{\rm DS}(T_h)}{\rho_\gamma(T_\gamma)},
\end{equation}    
where $\rho_{\rm DS}=\rho_{D}+\rho_{S}+\rho_{\phi}$. Here $\rho_D$ and $\rho_S$ are mass densities and are given by $\rho_i=m_i n_i$ ($i=D,S)$. The energy densities of the massless particles are
\begin{align}
\rho_\phi(T_h)&={3n_\phi T_h}, \\
\rho_\gamma(T_\gamma)&=\frac{\pi^2}{30}g_\gamma T^4_\gamma\,,
\end{align}
where $g_\gamma=2$. For the benchmarks of Table~\ref{tab1}, $\Delta N_{\rm eff}^{\rm BBN}$ ranges from $\mathcal{O}(10^{-4})$ to $\mathcal{O}(10^{-2})$ which is far too small and thus consistent with $N_{\rm eff}^{\rm SM}$. We emphasize  here that the reason for this is having a particle number density in the dark sector that never traces its equilibrium distribution owing to the small coupling between the two sectors. Since at BBN the only particles that are still relativistic are $\phi$ and $S$,  one can argue that increasing the couplings between $D$ and $(\phi, S)$ could increase the number densities of the latter causing a conflict with BBN. This, however, reduces the relic density of $D$. To bring it back up to its experimental value, an increase in $\delta$ is required, which is constrained by several experiments as shown in Fig.~\ref{fig5}. 

Near recombination, all particles have decoupled from the dark sector bath and $\phi$ is the only relativistic degree of freedom in the dark sector. Thus, the contribution to $N_{\rm eff}$ following the decay $S\to\phi\phi$ is given by
\begin{equation}
\Delta N_{\rm eff}=\frac{8}{7}\left(\frac{11}{4}\right)^{4/3}\frac{\rho_\phi(T_\phi)}{\rho_\gamma(T_\gamma)}.
\label{neff}
\end{equation}
We note here that an important property of $\phi$ are sufficiently large self-interactions. This is achieved via a four-point contact interaction $\phi\phi\to\phi\phi\propto\lambda_\phi^2$ and via the exchange of $S$ (in both $s$ and $t$ channels), where the latter is proportional to $\kappa_{\phi S}^2$, see Eq.~(\ref{potential}). Self-interactions prevent $\phi$ from free-streaming which can cause observable effects on the CMB power spectrum~\cite{Bashinsky:2003tk,Hou:2011ec}. An example in the SM are neutrinos which free-stream after their decoupling from photons. Such a free-streaming leads to a phase shift in the CMB and BAO peaks~\cite{Bashinsky:2003tk,Baumann:2015rya}, and the measurements from Planck are sensitive to any deviation in the evolution of perturbations~\cite{Baumann:2015rya,Friedland:2007vv,Follin:2015hya}. We can estimate the size of the coupling required to achieve efficient self-interactions of $\phi$ making it an ideal fluid. We will ignore the diagrams that are proportional to $\kappa_{\phi S}^2$ since they are very small (see Table~\ref{tab1}). The smallness of this coupling ensures that $S$ is long-lived and decays just prior to recombination. Thus, the only diagram remaining is the four-point contact interaction. The interaction rate is given by
\begin{equation}
n_\phi(T_\phi)\langle\sigma v\rangle_{\phi\phi\to\phi\phi}=\frac{9x\lambda^2_\phi}{64\pi^2}T_\phi,
\end{equation}  
where $x=n_\phi/n_\phi^{\rm eq}$. Comparing it to the Hubble rate, we can derive a lower bound on $\lambda_\phi$ such that
\begin{equation}
\lambda^2_\phi\geq \frac{64\pi^2}{9x}\left(\frac{8\pi^3}{90 M^2_{\rm Pl}}g_{\rm eff}\right)^{1/2}\xi^2 T_\gamma,
\end{equation}
where $g_{\rm eff}=g^v_{\rm eff}+g^h_{\rm eff}\xi^4$. For $T_\gamma\sim 1$ eV, $\xi\sim 0.5$ and $x\sim 10^{-2}$, we obtain $\lambda_\phi\gtrsim 10^{-12}$.  Thus even for such small values of the self-coupling one can achieve efficient self-interactions to prevent the free-streaming of $\phi$. 

\section{Reheating of the dark sector bath prior to recombination}\label{sec:reheating}

Contributions to $\Delta N_{\rm eff}$ at CMB have been considered before in the literature, but only in two contexts: (1) the visible and hidden sectors were initially in thermal contact and decouple later at a temperature $T_d$ causing the temperature of the hidden sector to vary as~\cite{Blinov:2020hmc}
\begin{equation}
\frac{T_\phi}{T_\gamma}=\left[\frac{h_{\rm eff}^v(T_\gamma)}{h_{\rm eff}^h(T_d)}\right]^{1/3},
\end{equation} 
and (2) the sectors were not in thermal equilibrium but eventually become thermalized~\cite{Berlin:2017ftj,Chacko:2003dt,Chacko:2004cz} where a hidden sector particle decays to neutrinos after it becomes non-relativistic~\cite{EscuderoAbenza:2020cmq,Escudero:2019gzq,Escudero:2021rfi}.
In the  model discussed here, the  particles in the dark sector are produced exclusively by SM processes. Furthermore, the massive particle $S$ and the relativistic species $\phi$ in the dark sector are produced in the right amount so as to satisfy BBN constraints while contributing effectively to 
$\Delta N_{\rm eff}^{\rm CMB}$ which can be obtained 
 entirely due to contributions from the 
dark sector and without participation of  the neutrino sector.

After decoupling of the dark sector species from the visible sector, 
the main coupled set of Boltzmann equations that we need to solve are
\begin{align}
\frac{d\rho_S}{dt}+3H\rho_S&=-\Gamma_S \,\rho_S, \\
\frac{d\rho_\phi}{dt}+4H\rho_\phi&=\Gamma_S \,\rho_S, \\
\frac{d\rho_D}{dt}+3H\rho_D&=0, \\
\frac{d\rho_\gamma}{dt}+4H\rho_\gamma&=0, 
\end{align}
where the Hubble parameter is given by
\begin{equation}
H(T)=\sqrt{\frac{8\pi}{3m^2_{\rm Pl}}(\rho_S+\rho_\phi+\rho_\gamma +\rho_D+3\rho_\nu)}. 
\end{equation}
Here, $m_{\rm Pl}=1.22\times 10^{19}$ GeV is the Planck mass and the decay width of $S\to\phi\phi$ is given by
\begin{equation}
\Gamma_{S}=\frac{\kappa_{\phi S}^2}{32\pi m_S}.
\end{equation}
Adding the Boltzmann equation for neutrinos (and for electrons) does not have any significant
impact on the analysis. We integrate the equations from the temperature at which the last dark sector particle decouples till the time of recombination, $T_\gamma\sim 1$ eV. The results for the evolution of the energy densities and $\Delta N_{\rm eff}$ with time are shown in Fig.~\ref{fig6}.

In the top panels of Fig.~\ref{fig6}, we exhibit the change in the energy densities as a function of time. One can see that the energy density of $\phi$ starts to increase as $S$ begins to decay. Eventually $\rho_S$ drops to zero and $\rho_\phi$ becomes closer to $\rho_\gamma$ which causes an increase in $\Delta N_{\rm eff}$ according to Eq.~(\ref{neff}). In the bottom panels, we show the variation in $\Delta N_{\rm eff}$ for four choices of the coupling $\lambda_{\phi S}$ (left) and $\kappa_{\phi S}$ (right). Here we see that $\Delta N_{\rm eff}^{\rm CMB}$ can be achieved in the required range of 0.2$-$0.5 for the chosen parameter set.
A smaller $\lambda_{\phi S}$ means the reaction $SS\to\phi\phi$ becomes less efficient leading to more $S$ particles available to decay to $\phi$ just before recombination which explains the increase in $\Delta N_{\rm eff}^{\rm CMB}$ for smaller  $\lambda_{\phi S}$. Furthermore, a smaller $\kappa_{\phi S}$ leads to the same result, i.e., $\Delta N_{\rm eff}^{\rm CMB}$ increases as the decay of $S$ happens closer to recombination.

\begin{figure}[H]
    \centering
  \includegraphics[width=0.495\textwidth]{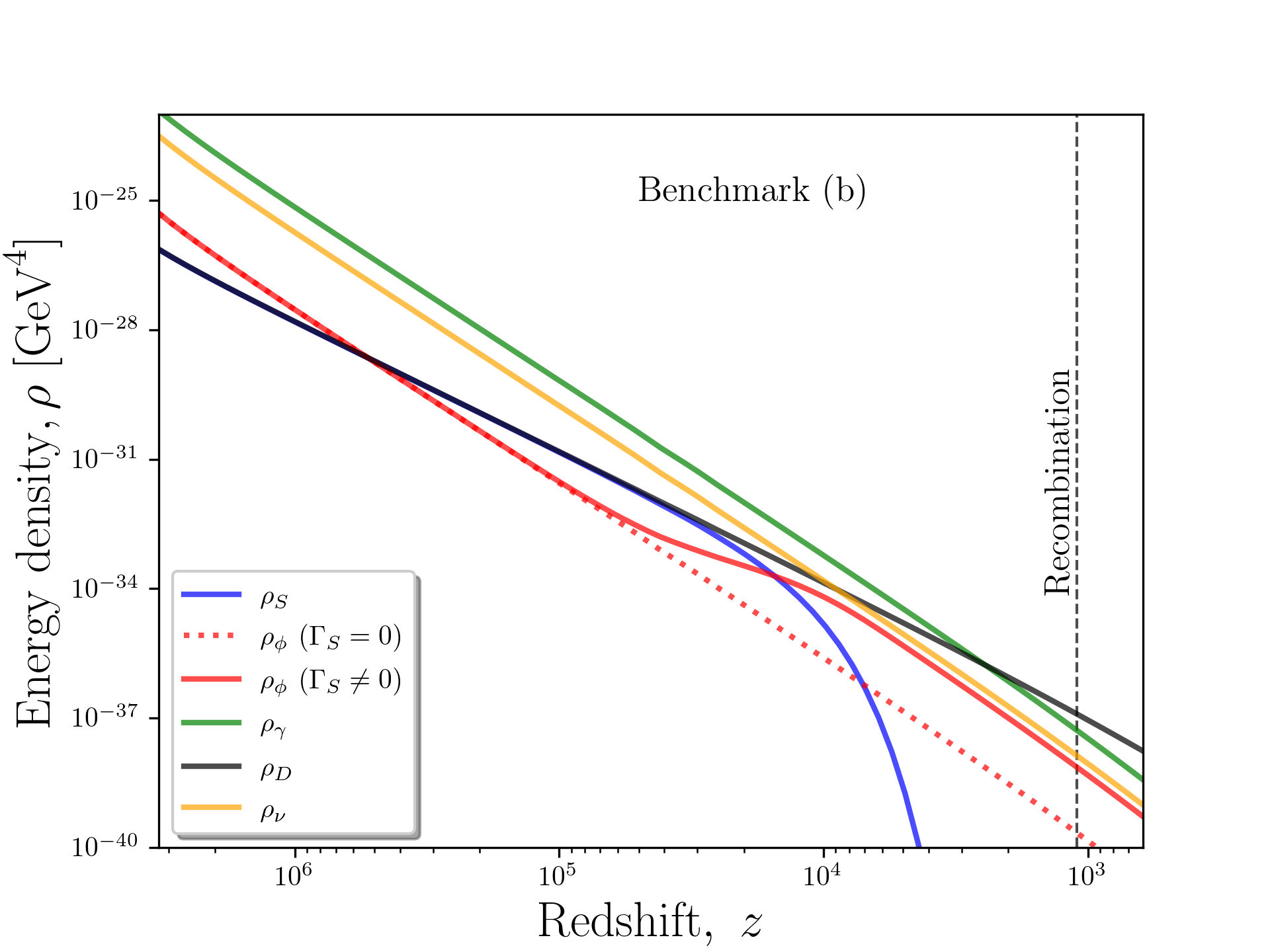}
  \includegraphics[width=0.495\textwidth]{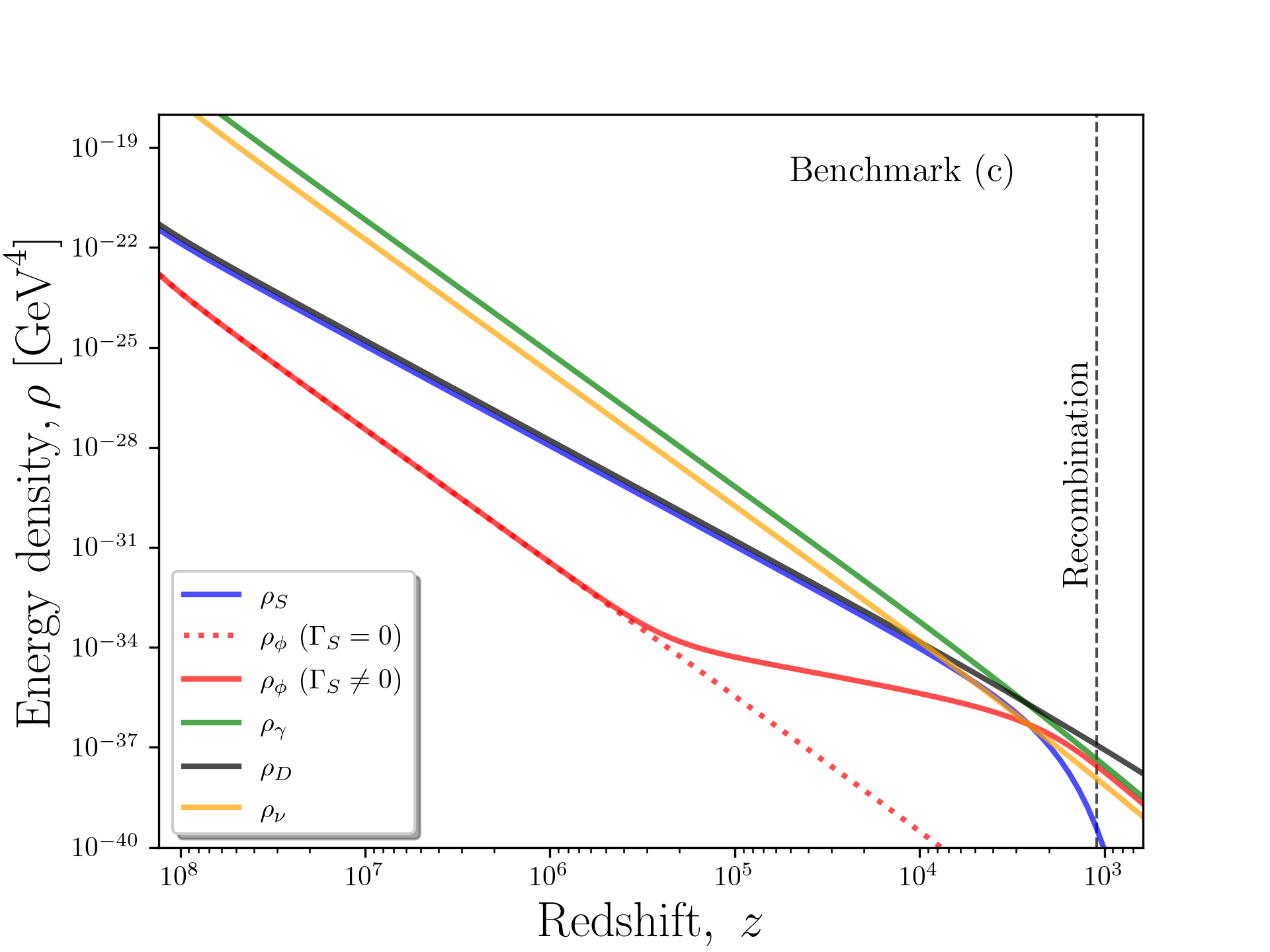} \\
  \includegraphics[width=0.495\textwidth]{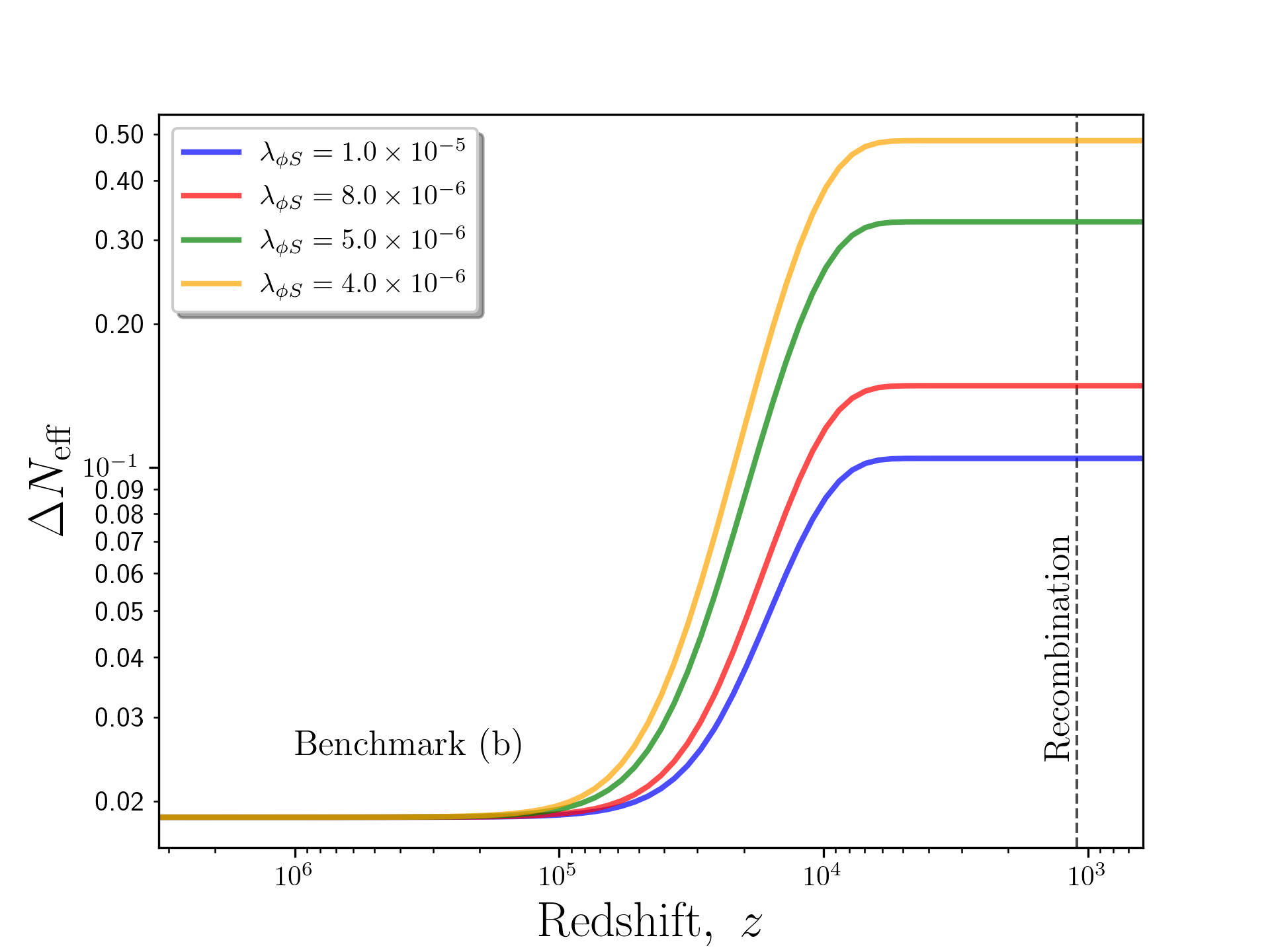}
  \includegraphics[width=0.495\textwidth]{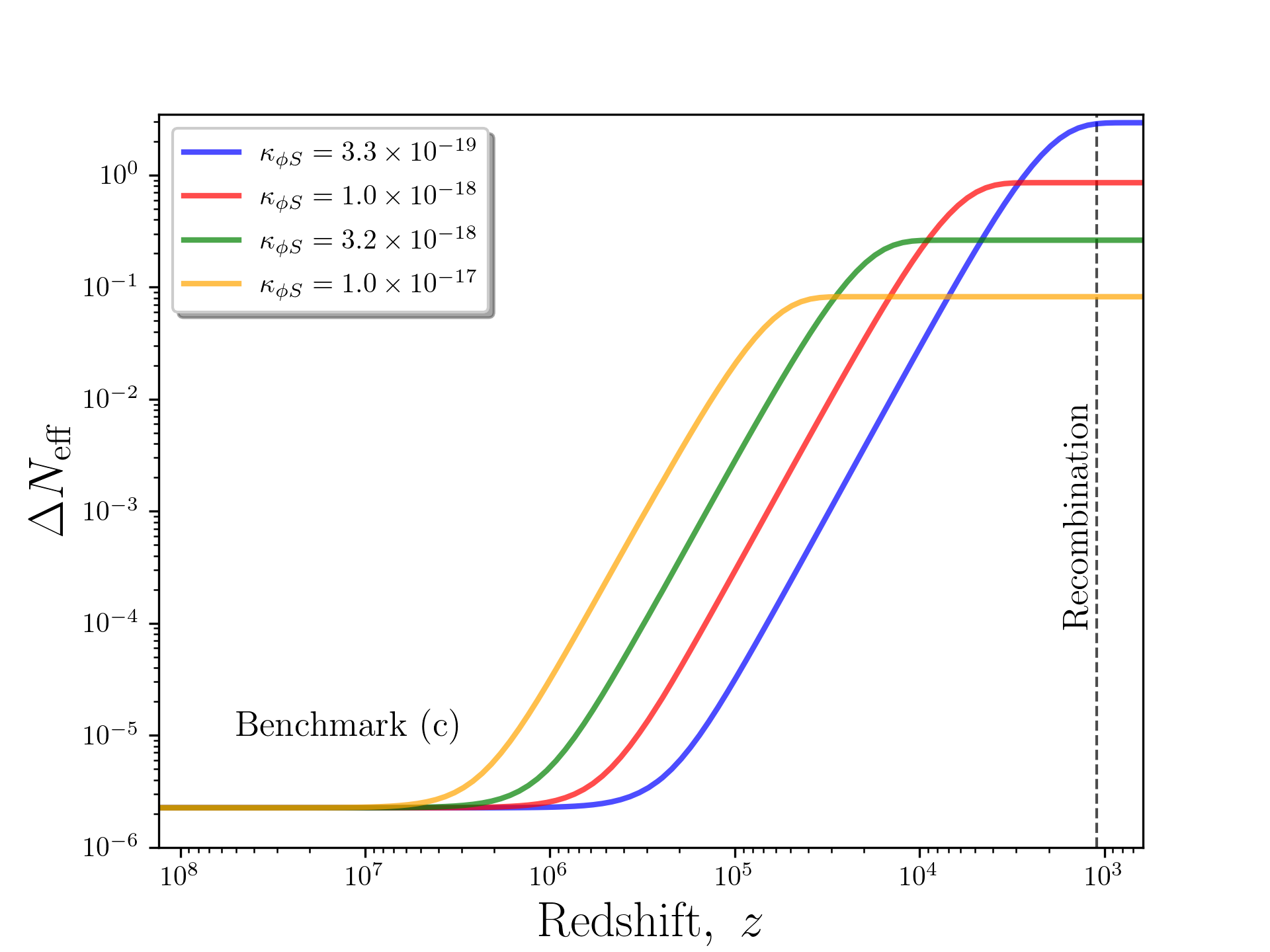}
    \caption{Top panels: Evolution of the energy densities of $S$, $\phi$, $D$, the neutrino and the photon with time for benchmark (b) (left) and benchmark (c) (right), where the dashed line shows $\rho_\phi$ in case $S$ does not decay. Bottom panels: Change in $\Delta N_{\rm eff}$ from decoupling to recombination for four values of the coupling $\lambda_{\phi S}$ (left) and for four values of the coupling $\kappa_{\phi S}$ (right). The other parameters are the same as for benchmarks (b) and (c) in Table~\ref{tab1}. } 
    \label{fig6}
\end{figure}

 We present in Fig.~\ref{fig7} an exclusion limit in the two parameters $y_S$ and $\lambda_{\phi S}$ for benchmark (b) in the left panel and benchmark (c) in the right panel. Similar limits can be realized for the other benchmarks. Here one finds
 that the parameter space for each $(m_D,m_{\gamma'})$ is constrained by the DM relic density and the requirement on early and late kinetic decoupling of dark sector particles. The region consistent with the DM relic density is shown in yellow and the one which gives $\Delta N_{\rm eff}^{\rm CMB}$ in the right range which may help alleviate the Hubble tension is shown in brown. Points in the overlap region are cosmologically consistent model points able to reduce the Hubble tension while in accordance with BBN constraints on $\Delta N_{\rm eff}$. 

\begin{figure}[H]
    \centering
\includegraphics[width=0.495\textwidth]{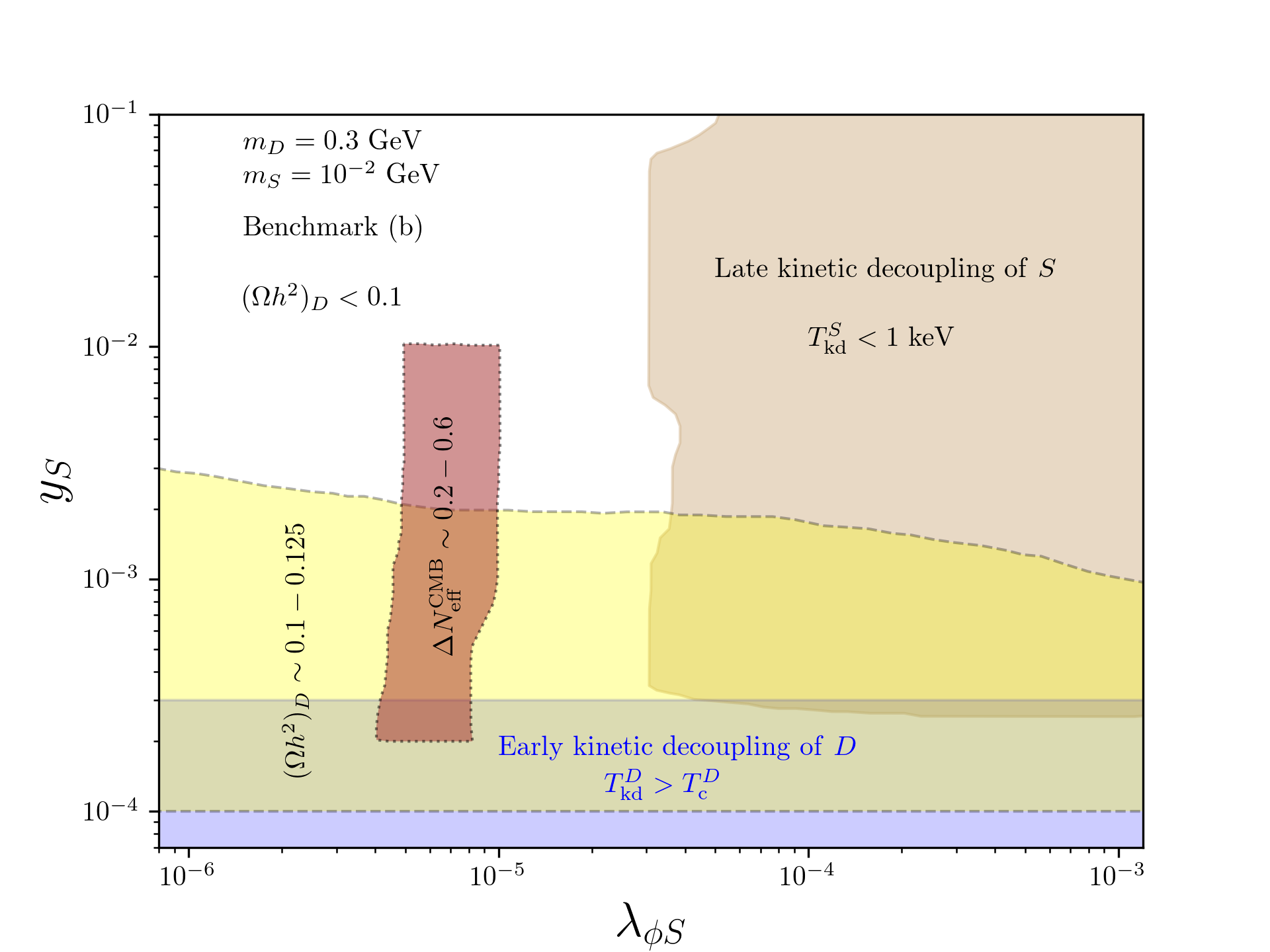}
\includegraphics[width=0.495\textwidth]{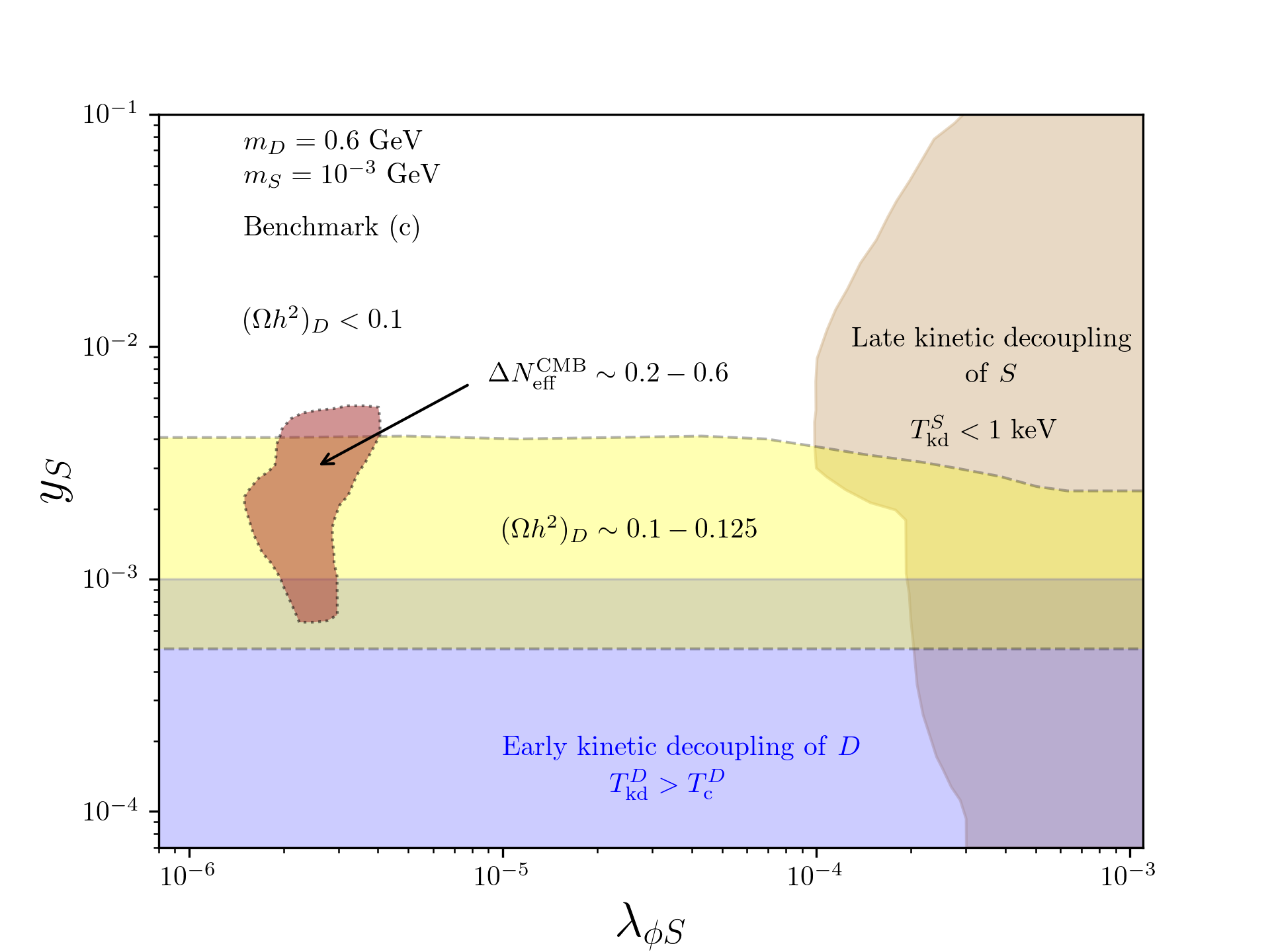}
    \caption{Exclusion limits in the $y_S$-$\lambda_{\phi S}$ plane for benchmark (b) (left panel) and benchmark (c) (right panel). The overlap between the brown and the yellow regions correspond to a parameter space consistent with all experiments.} 
    \label{fig7}
\end{figure}

Having extra relativistic degrees of freedom from a dark sector (i.e., a dark radiation) will have observable effects on the CMB power spectrum. The cosmological background is blind to the nature of the relativistic species when it comes to its effect on $H_0$. Both free-streaming and non-free-streaming dark radiation cause an increase in the expansion rate of the universe prior to matter-radiation equality. This additional radiation pushes more high-$\ell$ modes to enter the horizon during radiation-domination~\cite{Hu:2000ti} and since radiation-matter equality is required to occur at a fixed $z$~\cite{Hou:2011ec}, this creates a degeneracy between extra radiation and the matter density. As a result, the DM relic density is expected to grow.  Furthermore, changes in the comoving sound horizons at CMB last scattering $(r_*)$ and baryon drag $(r_d)$ are expected. Because of the stringent constraints from CMB on $r_*/D_M$ (which controls the angular scale of the CMB peaks), a change in the angular diameter distance $D_M$ should occur to keep that ratio fixed. Also, extra radiation is degenerate with the free electron fraction which is correlated with the Helium abundance that is successfully predicted by BBN. All of this shows how the different cosmological parameters are tightly related and to know the extent the additional particles $S$ and $\phi$ and their dynamics have on these parameters and especially on $H_0$ would require a more extensive analysis to fit the CMB data.  

\section{Conclusion\label{sec:conclu}}

The discrepancy between the high redshift ($z>1000$) and the low redshift ($z<1$) measurements
of the Hubble parameter, which now stands at $5\sigma$, is an important clue to the model underlying
the expansion of the universe. At early times, the Hubble parameter is extracted based on the
$\Lambda$CDM) model from measurements of the acoustic scale, which is precisely determined by Planck. Local direct measurements of the Hubble parameter $H_0$ have shown that this value is larger than the one calculated at early times, a tension that has recently grown to $5\sigma$. In order to help alleviate the tension, 
we propose a model which can deliver extra relativistic degrees of freedom at CMB times while respecting BBN bounds on $\Delta N_{\rm eff}$. 
Our analysis is based on a two-sector model, one visible and the other hidden, where we assume
a negligible particle abundance of the hidden sector particles in the early universe, and where the hidden sector
particles are generated by the out-of-equilibrium annihilation of SM particles into
 particles in the hidden sector. Because of that, the two sectors reside in heat baths at different temperatures with the hidden sector
 at a lower temperature than the visible sector. 
Thus our analysis uses a set of coupled Boltzmann equations which depend
on temperatures of both the visible and the hidden sectors and involves correlated temperature
evolution of both sectors. In this set up we show that for a range of kinetic mixing and dark photon mass consistent with experimental limits, the visible and the dark sectors do not thermalize which
 implies that the dark sector particles never reach their equilibrium number densities. However, their phase space distributions can still be described by a Maxwell-Boltzmann distribution since they achieve chemical and kinetic equilibrium during the dark sector thermal history. The fact that the dark  particles, especially the relativistic particle $\phi$, never reach their equilibrium distribution allows to evade the constraint on $\Delta N_{\rm eff}^{\rm BBN}$. With a specific choice of couplings, late decays of $S\to\phi\phi$ can contribute the right amount to $\Delta N_{\rm eff}^{\rm CMB}$ thus resolving the Hubble tension.  The model we present is cosmologically consistent since 
  the formalism allows for correlated evolution of the temperatures of the hidden and the visible
  sectors which are not in thermal equilibrium.  In fact it is the lack of thermal equilibrium between
  the hidden and visible sectors which forces consistency with the BBN constraint while allowing 
 the enhancement of $N_{\rm eff}$ at CMB  time.
 However, as noted earlier, while $\Delta N_{\rm eff}$ at recombination time
  would  weaken the  Hubble tension, a better understanding of how this affects the CMB power spectrum requires a fit of the cosmological parameters to the CMB data.

\noindent
{\bf Acknowledgments:}  
The research of AA and MK was supported by the BMBF under contract 05P21PMCAA and by the DFG through the Research Training Network 2149 ``Strong and Weak Interactions - from Hadrons to Dark Matter", while the research of PN was supported in part by the NSF Grant PHY-1913328.

\section{Further details of the analysis \label{sec:append}}
In this appendix we give further details of the analysis presented in the main body of the paper.
In Appendix~\ref{app:A} we give the formulae for the effective energy and entropy 
degrees of freedom, in  Appendix~\ref{app:B} we give details of  mixings of the two $U(1)$ sectors and in Appendix ~\ref{app:C} we give the Boltzmann equations with visible and hidden sector temperatures.

\appendix

\section{Degrees of freedom}{\label{app:A}
Here we focus on the hidden sector energy effective energy degrees of freedom $g^h_{\rm eff}$  in Eq.~(\ref{rho-1}) and on the hidden sector entropy effective degrees of freedom 
  $h^h_{\rm eff}$ in Eq.~(\ref{eq3.a}). They include degrees of freedom for the  dark photon, the dark fermion and the scalars $S$ and $\phi$ so that
\begin{align}
g^h_{\rm eff}&= g^{\gamma'}_{\rm eff} +\frac{7}{8}  g^D_{\rm eff}+g_\phi+g_S,~~~\text{and}~~~h^h_{\rm eff}= h^{\gamma'}_{\rm eff} + \frac{7}{8} h^D_{\rm eff}+h_\phi+h_S,
\end{align}
where $g_\phi=g_S=h_\phi=h_S=1$ and at  temperature $T_h$, $g_{\rm eff}$ and  $h_{\rm eff}$  for the particles $\gamma'$ and $D$ are 
given by~\cite{Hindmarsh:2005ix} 
\begin{equation}
\begin{aligned}
g^{\gamma'}_{\rm eff}& = \frac{45}{\pi^4} \int_{x_{\gamma'}}^{\infty} \frac{\sqrt{x^2-x_{\gamma'}^2} }{e^x-1 } x^2 dx,~~~\text{and}~~~h^{\gamma'}_{\rm eff}= \frac{45}{4\pi^4} \int_{x_{\gamma'}}^{\infty} \frac{\sqrt{x^2-x_{\gamma'}^2} }{e^x-1 } 
(4x^2-x_{\gamma'}^2) dx, \\
g^{D}_{\rm eff}& = \frac{60}{\pi^4} \int_{x_{D}}^{\infty} \frac{\sqrt{x^2-x_{D}^2} }{e^x+1 } x^2 dx,~~~\text{and}~~~h^{D}_{\rm eff}= \frac{15}{\pi^4} \int_{x_{D}}^{\infty} \frac{\sqrt{x^2-x_{D}^2} }{e^x+1 } 
(4x^2-x_D^2) dx.
\end{aligned}
\label{hdof}
\end{equation}
Here $x_{\gamma'}$ and $x_D$ are defined so that $x_{\gamma'}= m_{\gamma'}/T_h$ and $x_D= m_D/T_h$. 
The  limit 
$x_{\gamma'}\to 0$  gives $g^{\gamma'}_{\rm eff}= h^{\gamma'}_{\rm eff}\to 3$ and the limit $x_D\to 0$ gives $g^{D}_{\rm eff}= h^{D}_{\rm eff}\to 4$. Further details of the model are given in Appendix~\ref{app:B}.

\section{Model details}\label{app:B}

In this appendix we give further details of the extended model with an additional $U(1)_X$ 
gauge group factor beyond the SM gauge group. In addition to the electroweak SM gauge fields  
$A^\mu_a$ ($a=$ 1$-$3) of $SU(2)_L$ and $B_\mu$ of $U(1)_Y$, we have a $U(1)_X$
gauge field $C_\mu$. The mixing of the gauge fields in the neutral sector leads to a $3\times 3$ mass
square matrix whose diagonalization leads to the photon, the $Z$ boson and a new massive neutral
field which we label as the dark photon. 
Here we display the form of the couplings that enter $\Delta\mathcal{L}^{\rm int}$. Thus the  couplings $g_{Z}^D , g_{\gamma}^D$ and $g^D_{\gamma'}$ that appear  in   $\Delta\mathcal{L}^{\rm int}$ (see Eq.~(\ref{D-darkphoton}))
are given by 
\begin{align}
g_Z^D &= g_X Q_X (\mathcal{R}_{12}- s_{\delta} \mathcal{R}_{22}), 
  ~g_{\gamma}^D &= g_X Q_X (\mathcal{R}_{13}- s_{\delta} \mathcal{R}_{23}), 
 ~ g_{\gamma'}^D&= g_X Q_X (\mathcal{R}_{11}- s_{\delta} \mathcal{R}_{21}),
 \end{align}
where the matrix $\mathcal{R}$ is given  by Eq.~(23) of~\cite{Feldman:2007wj}. It involves three Euler angles $(\theta, \phi, \psi)$ 
 which diagonalize the Stueckelberg mass  matrix such that $\mathcal{R}^T\mathcal{M}^2 \mathcal {R}= \text{diag}(m^2_{\gamma'},  m_Z^2, 0)$,
 where $\mathcal{M}^2$ is defined by Eq.~(21) of~\cite{Feldman:2007wj}.
The angles $\phi, \theta, \psi$ are defined as
\begin{equation}
 \tan\phi=-s_{\delta}, ~~~ \tan\theta=\frac{g_Y}{g_2}c_{\delta}\cos\phi,
~\tan2\psi=\frac{2\bar\epsilon m^2_Z\sin\theta}{m^2_{\gamma'}-m^2_Z+(m^2_{\gamma'}+m^2_Z-m^2_W)\bar\epsilon^2},
 \label{hid-exact}
\end{equation}
 where $s_{\delta}=\sinh\delta$ and $c_{\delta} = \cosh\delta$.

 \section{Boltzmann equations with visible and hidden sector temperatures}\label{app:C}   

Since, in general, the visible and the hidden sector reside in different temperature
   baths, the Boltzmann equations  for the yields $Y_D$, $Y_{\gamma'}$, $Y_\phi$ and $Y_S$ 
   involve these two temperatures and are given by
\begin{align}
\label{yphi}
\frac{dY_{\phi}}{dT_h}=&-\frac{\mathbb{s}}{H}\left(\frac{d\rho_h/dT_h}{4\zeta\rho_h-j_h/H}\right)\Bigg[\frac{1}{2}\langle\sigma v\rangle_{D\bar{D}\to\phi\gamma'}(T_h)\left(Y^2_D-Y_D^{\rm eq}(T_h)^2\frac{Y_{\phi}Y_{\gamma'}}{Y^{\rm eq}_{\phi}(T_h)Y^{\rm eq}_{\gamma'}(T_h)}\right) \nonumber \\
&\hspace{4cm}+\frac{1}{2}\langle\sigma v\rangle_{D\bar{D}\to\phi S}(T_h)\left(Y^2_D-Y_D^{\rm eq}(T_h)^2\frac{Y_{\phi}Y_{S}}{Y^{\rm eq}_{\phi}(T_h)Y^{\rm eq}_{S}(T_h)}\right) \nonumber \\
&\hspace{4cm}+\frac{1}{2}\langle\sigma v\rangle_{D\bar{D}\to\phi\phi}(T_h)\left(Y^2_D-Y_D^{\rm eq}(T_h)^2\frac{Y_{\phi}^2}{Y^{\rm eq}_{\phi}(T_h)^2}\right) \nonumber \\
&\hspace{4cm}-\langle\sigma v\rangle_{\phi\phi\to SS}(T_h)\left(Y^2_\phi-Y_\phi^{\rm eq}(T_h)^2\frac{Y_{S}^2}{Y^{\rm eq}_{S}(T_h)^2}\right) \nonumber \\
&\hspace{4cm}-\langle\sigma v\rangle_{\phi\phi\to S}(T_h)\left(Y^2_\phi-Y_\phi^{\rm eq}(T_h)^2\frac{Y_{S}}{Y^{\rm eq}_{S}(T_h)}\right) \Bigg], 
\end{align}

\begin{align}
\label{yS}
\frac{dY_{S}}{dT_h}=&-\frac{\mathbb{s}}{H}\left(\frac{d\rho_h/dT_h}{4\zeta\rho_h-j_h/H}\right)\Bigg[\frac{1}{2}\langle\sigma v\rangle_{D\bar{D}\to S\gamma'}(T_h)\left(Y^2_D-Y_D^{\rm eq}(T_h)^2\frac{Y_{S}Y_{\gamma'}}{Y^{\rm eq}_{S}(T_h)Y^{\rm eq}_{\gamma'}(T_h)}\right) \nonumber \\
&\hspace{4cm}+\frac{1}{2}\langle\sigma v\rangle_{D\bar{D}\to\phi S}(T_h)\left(Y^2_D-Y_D^{\rm eq}(T_h)^2\frac{Y_{\phi}Y_{S}}{Y^{\rm eq}_{\phi}(T_h)Y^{\rm eq}_{S}(T_h)}\right) \nonumber \\
&\hspace{4cm}+\frac{1}{2}\langle\sigma v\rangle_{D\bar{D}\to SS}(T_h)\left(Y^2_D-Y_D^{\rm eq}(T_h)^2\frac{Y_{S}^2}{Y^{\rm eq}_{S}(T_h)^2}\right) \nonumber \\
&\hspace{4cm}+\langle\sigma v\rangle_{\phi\phi\to SS}(T_h)\left(Y^2_\phi-Y_\phi^{\rm eq}(T_h)^2\frac{Y_{S}^2}{Y^{\rm eq}_{S}(T_h)^2}\right) \nonumber \\
&\hspace{4cm}+\langle\sigma v\rangle_{\phi\phi\to S}(T_h)\left(Y^2_\phi-Y_\phi^{\rm eq}(T_h)^2\frac{Y_{S}}{Y^{\rm eq}_{S}(T_h)}\right) \Bigg], 
\end{align}

\begin{align}
\label{yg}
\frac{dY_{\gamma'}}{dT_h}=&-\frac{\mathbb{s}}{H}\left(\frac{d\rho_h/dT_h}{4\zeta\rho_h-j_h/H}\right)\Bigg[-\langle\sigma v\rangle_{\gamma'\gamma'\to D\bar{D}}(T_h)\left(Y^2_{\gamma'}-Y^{\rm eq}_{\gamma'}(T_h)^2\frac{Y^2_D}{Y^{\rm eq}_D(T_h)^2}\right) \nonumber \\
&\hspace{4cm}+\frac{1}{2}\langle\sigma v\rangle_{D\bar{D}\to\phi\gamma'}(T_h)\left(Y^2_D-Y_D^{\rm eq}(T_h)^2\frac{Y_{\phi}Y_{\gamma'}}{Y^{\rm eq}_{\phi}(T_h)Y^{\rm eq}_{\gamma'}(T_h)}\right) \nonumber \\
&\hspace{4cm}+\frac{1}{2}\langle\sigma v\rangle_{D\bar{D}\to S\gamma'}(T_h)\left(Y^2_D-Y_D^{\rm eq}(T_h)^2\frac{Y_{S}Y_{\gamma'}}{Y^{\rm eq}_{S}(T_h)Y^{\rm eq}_{\gamma'}(T_h)}\right) \nonumber \\
&\hspace{4cm}-\frac{1}{\mathbb{s}}\langle\Gamma_{\gamma'\to i\bar{i}}\rangle(T_h)\,Y_{\gamma'}+\langle\sigma v\rangle_{i\bar{i}\to\gamma'}(T)Y_i^{\rm eq}(T)^2 \nonumber \\
&\hspace{4cm}-\frac{1}{\mathbb{s}}\langle\Gamma_{\gamma'\to D\bar{D}}\rangle(T_h)\left(Y_{\gamma'}-Y^{\rm eq}_{\gamma'}(T_h)\frac{Y^2_D}{Y^{\rm eq}_D(T_h)^2}\right)\Bigg],
\end{align}

\begin{align}
\label{yD}
\frac{dY_D}{dT_h}=&-\frac{\mathbb{s}}{H}\Big(\frac{d\rho_h/dT_h}{4\zeta\rho_h-j_h/H}\Big)\Bigg[\langle\sigma v\rangle_{i\bar{i}\to D \bar D}(T) Y_D^{\rm eq}(T)^2 \nonumber \\
&\hspace{4cm}-\frac{1}{2}\langle\sigma v\rangle_{D\bar{D}\to\gamma'\gamma'}(T_h)\left(Y^2_D-Y_D^{\rm eq}(T_h)^2\frac{Y^2_{\gamma'}}{Y^{\rm eq}_{\gamma'}(T_h)^2}\right) \nonumber\\
&\hspace{4cm}-\frac{1}{2}\langle\sigma v\rangle_{D\bar{D}\to\phi\phi}(T_h)\left(Y^2_D-Y_D^{\rm eq}(T_h)^2\frac{Y_{\phi}^2}{Y^{\rm eq}_{\phi}(T_h)^2}\right) \nonumber \\
&\hspace{4cm}-\frac{1}{2}\langle\sigma v\rangle_{D\bar{D}\to SS}(T_h)\left(Y^2_D-Y_D^{\rm eq}(T_h)^2\frac{Y_{S}^2}{Y^{\rm eq}_{S}(T_h)^2}\right) \nonumber \\
&\hspace{4cm}-\frac{1}{2}\langle\sigma v\rangle_{D\bar{D}\to\phi\gamma'}(T_h)\left(Y^2_D-Y_D^{\rm eq}(T_h)^2\frac{Y_{\phi}Y_{\gamma'}}{Y^{\rm eq}_{\phi}(T_h)Y^{\rm eq}_{\gamma'}(T_h)}\right) \nonumber \\
&\hspace{4cm}-\frac{1}{2}\langle\sigma v\rangle_{D\bar{D}\to S\gamma'}(T_h)\left(Y^2_D-Y_D^{\rm eq}(T_h)^2\frac{Y_{S}Y_{\gamma'}}{Y^{\rm eq}_{S}(T_h)Y^{\rm eq}_{\gamma'}(T_h)}\right) \nonumber \\
&\hspace{4cm}-\frac{1}{2}\langle\sigma v\rangle_{D\bar{D}\to S\phi}(T_h)\left(Y^2_D-Y_D^{\rm eq}(T_h)^2\frac{Y_{S}Y_{\phi}}{Y^{\rm eq}_{S}(T_h)Y^{\rm eq}_{\phi}(T_h)}\right) \nonumber \\
&\hspace{4cm}-\frac{1}{2}\langle\sigma v\rangle_{D\bar{D}\to\gamma'}(T_h)\,Y^2_D+\frac{1}{\mathbb{s}}\langle\Gamma_{\gamma'\to D\bar{D}}\rangle(T_h)\,Y_{\gamma'}\Bigg].
\end{align}
The thermally averaged cross sections appearing in Eqs.~(\ref{yphi})$-$(\ref{yD}) are given by
\begin{equation}
\langle\sigma v\rangle_{a\bar{a}\to bc}(T)=\frac{1}{8 m^4_a T K^2_2(m_a/T)}
\int_{4m_a^2}^{\infty} ds ~\sigma(s) \sqrt{s}\, (s-4m_a^2)K_1(\sqrt{s}/T),
\end{equation}
and
\begin{align}
n_i^{\rm eq}(T)^2\langle\sigma v\rangle_{i\bar{i}\to\gamma'}(T)&= 
\frac{T}{32\pi^4}\int_{s_0}^{\infty} ds ~\sigma(s) \sqrt{s}\, (s-s_0)K_1(\sqrt{s}/T),
\end{align}
where $K_1$ is the modified Bessel function of the second kind and degree one and $s_0$ is the minimum of the Mandelstam variable $s$. Note the appearance of the prefactor $\frac{\mathbb{s}}{H}\Big(\frac{d\rho_h/dT_h}{4\zeta\rho_h-j_h/H}\Big)$ in the Boltzmann equations which has a dependence on the source $j_h$. 
  
The equilibrium yield is given by
\begin{equation}
Y^{\rm eq}_i=\frac{n_i^{\rm eq}}{\mathbb{s}}=\frac{g_i}{2\pi^2 \mathbb{s}}m_i^2 T K_2(m_i/T),
\end{equation}
where $n_i$ is the number density of a SM particle species $i$,  
$g_i$ is its number of degrees of freedom, and $m_i$ its mass. Further,  $K_2$ is the modified Bessel function of the second kind and degree two and the $J$ functions  
in Eq.~(\ref{jh}) are given by
\begin{align}
n^{\rm eq}_i(T)^2 J(i~\bar{i}\to D\bar{D})(T)&=\frac{T}{32\pi^4}\int_{s_0}^{\infty}ds~\sigma_{D\bar{D}\to i\bar{i}}s(s-s_0)K_2(\sqrt{s}/T), \\
n^{\rm eq}_i(T)^2 J(i~\bar{i}\to \gamma')(T)&=\frac{T}{32\pi^4}\int_{s_0}^{\infty}ds~\sigma_{i\bar{i}\to \gamma'}s(s-s_0)K_2(\sqrt{s}/T), \\
n_{\gamma'}J(\gamma'\to f\bar{f})(T_h)&=n_{\gamma'}m_{\gamma'}\Gamma_{\gamma'\to f\bar{f}}, \\
{J(ab\to cd)(T_h)}&{=\frac{1}{8T_h m_a^2 m_b^2 K_2(m_a/T_h) K_2(m_b/T_h)} }\nonumber \\
& {\times\int_{(m_a+m_b)^2}^{\infty} ds~\sigma_{ab\to cd}(s) s[s-(m_a+m_b)^2]K_2(\sqrt{s}/T_h).}
\end{align}
We note in passing that in 
   Eq.~(\ref{yg}) the processes $i~\bar{i}\to \gamma' \gamma,
\gamma'Z, \gamma'\gamma'$ also contribute. However, their contributions are relatively small compared to those of $i~\bar{i}\to \gamma'$.


\begin{thebibliography}{999}

\bibitem{Riess:2021jrx}
A.~G.~Riess, W.~Yuan, L.~M.~Macri, D.~Scolnic, D.~Brout, S.~Casertano, D.~O.~Jones, Y.~Murakami, L.~Breuval and T.~G.~Brink, \textit{et al.}
[arXiv:2112.04510 [astro-ph.CO]].

\bibitem{Planck:2018vyg}
N.~Aghanim \textit{et al.} [Planck],
Astron. Astrophys. \textbf{641}, A6 (2020)
[erratum: Astron. Astrophys. \textbf{652}, C4 (2021)]
doi:10.1051/0004-6361/201833910
[arXiv:1807.06209 [astro-ph.CO]].

\bibitem{DiValentino:2021izs}
E.~Di Valentino, O.~Mena, S.~Pan, L.~Visinelli, W.~Yang, A.~Melchiorri, D.~F.~Mota, A.~G.~Riess and J.~Silk,
Class. Quant. Grav. \textbf{38}, no.15, 153001 (2021)
doi:10.1088/1361-6382/ac086d
[arXiv:2103.01183 [astro-ph.CO]].

\bibitem{Solomon:2022qqf}
R.~Solomon, G.~Agarwal and D.~Stojkovic,
[arXiv:2201.03127 [hep-ph]].

\bibitem{Fernandez-Martinez:2021ypo}
E.~Fernandez-Martinez, M.~Pierre, E.~Pinsard and S.~Rosauro-Alcaraz,
Eur. Phys. J. C \textbf{81}, no.10, 954 (2021)
doi:10.1140/epjc/s10052-021-09760-y
[arXiv:2106.05298 [hep-ph]].

\bibitem{Escudero:2021rfi}
M.~Escudero and S.~J.~Witte,
Eur. Phys. J. C \textbf{81}, no.6, 515 (2021)
doi:10.1140/epjc/s10052-021-09276-5
[arXiv:2103.03249 [hep-ph]].

\bibitem{Gehrlein:2019iwl}
J.~Gehrlein and M.~Pierre,
JHEP \textbf{02}, 068 (2020)
doi:10.1007/JHEP02(2020)068
[arXiv:1912.06661 [hep-ph]].

\bibitem{Escudero:2019gzq}
M.~Escudero, D.~Hooper, G.~Krnjaic and M.~Pierre,
JHEP \textbf{03}, 071 (2019)
doi:10.1007/JHEP03(2019)071
[arXiv:1901.02010 [hep-ph]].

\bibitem{Cyburt:2015mya}
R.~H.~Cyburt, B.~D.~Fields, K.~A.~Olive and T.~H.~Yeh,
Rev. Mod. Phys. \textbf{88}, 015004 (2016)
doi:10.1103/RevModPhys.88.015004
[arXiv:1505.01076 [astro-ph.CO]].

\bibitem{Pitrou:2018cgg}
C.~Pitrou, A.~Coc, J.~P.~Uzan and E.~Vangioni,
Phys. Rept. \textbf{754}, 1-66 (2018)
doi:10.1016/j.physrep.2018.04.005
[arXiv:1801.08023 [astro-ph.CO]].

\bibitem{Mangano:2005cc}
G.~Mangano, G.~Miele, S.~Pastor, T.~Pinto, O.~Pisanti and P.~D.~Serpico,
Nucl. Phys. B \textbf{729}, 221-234 (2005)
doi:10.1016/j.nuclphysb.2005.09.041
[arXiv:hep-ph/0506164 [hep-ph]].

\bibitem{Riess:2018uxu}
A.~G.~Riess, S.~Casertano, W.~Yuan, L.~Macri, J.~Anderson, J.~W.~MacKenty, J.~Bradley Bowers, K.~I.~Clubb, A.~V.~Filippenko and D.~O.~Jones, \textit{et al.}
Astrophys. J. \textbf{855}, no.2, 136 (2018)
doi:10.3847/1538-4357/aaadb7
[arXiv:1801.01120 [astro-ph.SR]].

\bibitem{Riess:2019cxk}
A.~G.~Riess, S.~Casertano, W.~Yuan, L.~M.~Macri and D.~Scolnic,
Astrophys. J. \textbf{876}, no.1, 85 (2019)
doi:10.3847/1538-4357/ab1422
[arXiv:1903.07603 [astro-ph.CO]].

\bibitem{Pan-STARRS1:2017jku}
D.~M.~Scolnic \textit{et al.} [Pan-STARRS1],
Astrophys. J. \textbf{859}, no.2, 101 (2018)
doi:10.3847/1538-4357/aab9bb
[arXiv:1710.00845 [astro-ph.CO]].

\bibitem{Seto:2021xua}
O.~Seto and Y.~Toda,
Phys. Rev. D \textbf{103}, no.12, 123501 (2021)
doi:10.1103/PhysRevD.103.123501
[arXiv:2101.03740 [astro-ph.CO]].

\bibitem{Seto:2021tad}
O.~Seto and Y.~Toda,
Phys. Rev. D \textbf{104}, no.6, 063019 (2021)
doi:10.1103/PhysRevD.104.063019
[arXiv:2104.04381 [astro-ph.CO]].

\bibitem{Vagnozzi:2019ezj}
S.~Vagnozzi,
Phys. Rev. D \textbf{102}, no.2, 023518 (2020)
doi:10.1103/PhysRevD.102.023518
[arXiv:1907.07569 [astro-ph.CO]].

\bibitem{Dainotti:2021pqg}
M.~G.~Dainotti, B.~De Simone, T.~Schiavone, G.~Montani, E.~Rinaldi and G.~Lambiase,
Astrophys. J. \textbf{912}, no.2, 150 (2021)
doi:10.3847/1538-4357/abeb73
[arXiv:2103.02117 [astro-ph.CO]].

\bibitem{Dainotti:2022bzg}
M.~G.~Dainotti, B.~De Simone, T.~Schiavone, G.~Montani, E.~Rinaldi, G.~Lambiase, M.~Bogdan and S.~Ugale,
Galaxies \textbf{10}, no.1, 24 (2022)
doi:10.3390/galaxies10010024
[arXiv:2201.09848 [astro-ph.CO]].

\bibitem{Holdom:1985ag}
B.~Holdom,
Phys. Lett. B \textbf{166}, 196-198 (1986)
doi:10.1016/0370-2693(86)91377-8

\bibitem{Holdom:1990xp}
B.~Holdom,
Phys. Lett. B \textbf{259}, 329-334 (1991)
doi:10.1016/0370-2693(91)90836-F

\bibitem{Kors:2004dx}
B.~Kors and P.~Nath,
Phys. Lett. B \textbf{586}, 366-372 (2004)
doi:10.1016/j.physletb.2004.02.051
[arXiv:hep-ph/0402047 [hep-ph]].

\bibitem{Kors:2004ri}
B.~Kors and P.~Nath,
JHEP \textbf{12}, 005 (2004)
doi:10.1088/1126-6708/2004/12/005
[arXiv:hep-ph/0406167 [hep-ph]].

\bibitem{Cheung:2007ut}
K.~Cheung and T.~C.~Yuan,
JHEP \textbf{03}, 120 (2007)
doi:10.1088/1126-6708/2007/03/120
[arXiv:hep-ph/0701107 [hep-ph]].

\bibitem{Feldman:2007wj}
D.~Feldman, Z.~Liu and P.~Nath,
Phys. Rev. D \textbf{75}, 115001 (2007)
doi:10.1103/PhysRevD.75.115001
[arXiv:hep-ph/0702123 [hep-ph]].

\bibitem{Aboubrahim:2020lnr}
A.~Aboubrahim, W.~Z.~Feng, P.~Nath and Z.~Y.~Wang,
Phys. Rev. D \textbf{103}, no.7, 075014 (2021)
doi:10.1103/PhysRevD.103.075014
[arXiv:2008.00529 [hep-ph]].

\bibitem{Aboubrahim:2020afx}
A.~Aboubrahim, T.~Ibrahim, M.~Klasen and P.~Nath,
Eur. Phys. J. C \textbf{81}, no.8, 680 (2021)
doi:10.1140/epjc/s10052-021-09483-0
[arXiv:2012.10795 [hep-ph]].

\bibitem{Aboubrahim:2021ycj}
A.~Aboubrahim, W.~Z.~Feng, P.~Nath and Z.~Y.~Wang,
JHEP \textbf{06}, 086 (2021)
doi:10.1007/JHEP06(2021)086
[arXiv:2103.15769 [hep-ph]].

\bibitem{Aboubrahim:2021ohe}
A.~Aboubrahim, P.~Nath and Z.~Y.~Wang,
JHEP \textbf{12}, 148 (2021)
doi:10.1007/JHEP12(2021)148
[arXiv:2108.05819 [hep-ph]].

\bibitem{Aboubrahim:2021dei}
A.~Aboubrahim, W.~Z.~Feng, P.~Nath and Z.~Y.~Wang,
[arXiv:2106.06494 [hep-ph]].

\bibitem{Foot:2014uba}
R.~Foot and S.~Vagnozzi,
Phys. Rev. D \textbf{91}, 023512 (2015)
doi:10.1103/PhysRevD.91.023512
[arXiv:1409.7174 [hep-ph]].

\bibitem{Foot:2016wvj}
R.~Foot and S.~Vagnozzi,
JCAP \textbf{07}, 013 (2016)
doi:10.1088/1475-7516/2016/07/013
[arXiv:1602.02467 [astro-ph.CO]].

\bibitem{Drees:2015exa}
M.~Drees, F.~Hajkarim and E.~R.~Schmitz,
JCAP \textbf{06}, 025 (2015)
doi:10.1088/1475-7516/2015/06/025
[arXiv:1503.03513 [hep-ph]].

\bibitem{Aghanim:2018eyx}
N.~Aghanim \textit{et al.} [Planck],
Astron. Astrophys. \textbf{641}, A6 (2020)
[erratum: Astron. Astrophys. \textbf{652}, C4 (2021)]
doi:10.1051/0004-6361/201833910
[arXiv:1807.06209 [astro-ph.CO]].

\bibitem{Binder:2017rgn}
T.~Binder, T.~Bringmann, M.~Gustafsson and A.~Hryczuk,
Phys. Rev. D \textbf{96}, no.11, 115010 (2017)
[erratum: Phys. Rev. D \textbf{101}, no.9, 099901 (2020)]
doi:10.1103/PhysRevD.96.115010
[arXiv:1706.07433 [astro-ph.CO]].

\bibitem{Binder:2021bmg}
T.~Binder, T.~Bringmann, M.~Gustafsson and A.~Hryczuk,
Eur. Phys. J. C \textbf{81}, no.7, 577 (2021)
doi:10.1140/epjc/s10052-021-09357-5
[arXiv:2103.01944 [hep-ph]].

\bibitem{Bringmann:2009vf}
T.~Bringmann,
New J. Phys. \textbf{11}, 105027 (2009)
doi:10.1088/1367-2630/11/10/105027
[arXiv:0903.0189 [astro-ph.CO]].

\bibitem{Bringmann:2006mu}
T.~Bringmann and S.~Hofmann,
JCAP \textbf{04}, 016 (2007)
[erratum: JCAP \textbf{03}, E02 (2016)]
doi:10.1088/1475-7516/2007/04/016
[arXiv:hep-ph/0612238 [hep-ph]].

\bibitem{Gondolo:2012vh}
P.~Gondolo, J.~Hisano and K.~Kadota,
Phys. Rev. D \textbf{86}, 083523 (2012)
doi:10.1103/PhysRevD.86.083523
[arXiv:1205.1914 [hep-ph]].

\bibitem{Bringmann:2016ilk}
T.~Bringmann, H.~T.~Ihle, J.~Kersten and P.~Walia,
Phys. Rev. D \textbf{94}, no.10, 103529 (2016)
doi:10.1103/PhysRevD.94.103529
[arXiv:1603.04884 [hep-ph]].

\bibitem{Bergsma:1985qz}
F.~Bergsma \textit{et al.} [CHARM],
Phys. Lett. B \textbf{157}, 458-462 (1985)
doi:10.1016/0370-2693(85)90400-9

\bibitem{Tsai:2019mtm}
Y.~D.~Tsai, P.~deNiverville and M.~X.~Liu,
Phys. Rev. Lett. \textbf{126}, no.18, 181801 (2021)
doi:10.1103/PhysRevLett.126.181801
[arXiv:1908.07525 [hep-ph]].

\bibitem{Riordan:1987aw}
E.~M.~Riordan, M.~W.~Krasny, K.~Lang, P.~De Barbaro, A.~Bodek, S.~Dasu, N.~Varelas, X.~Wang, R.~G.~Arnold and D.~Benton, \textit{et al.}
Phys. Rev. Lett. \textbf{59}, 755 (1987)
doi:10.1103/PhysRevLett.59.755

\bibitem{Blumlein:1990ay}
J.~Blumlein, J.~Brunner, H.~J.~Grabosch, P.~Lanius, S.~Nowak, C.~Rethfeldt, H.~E.~Ryseck, M.~Walter, D.~Kiss and Z.~Jaki, \textit{et al.}
Z. Phys. C \textbf{51}, 341-350 (1991)
doi:10.1007/BF01548556

\bibitem{Blumlein:1991xh}
J.~Blumlein, J.~Brunner, H.~J.~Grabosch, P.~Lanius, S.~Nowak, C.~Rethfeldt, H.~E.~Ryseck, M.~Walter, D.~Kiss and Z.~Jaki, \textit{et al.}
Int. J. Mod. Phys. A \textbf{7}, 3835-3850 (1992)
doi:10.1142/S0217751X9200171X

\bibitem{Blumlein:2013cua}
J.~Bl\"umlein and J.~Brunner,
Phys. Lett. B \textbf{731}, 320-326 (2014)
doi:10.1016/j.physletb.2014.02.029
[arXiv:1311.3870 [hep-ph]].

\bibitem{Blumlein:2011mv}
J.~Blumlein and J.~Brunner,
Phys. Lett. B \textbf{701}, 155-159 (2011)
doi:10.1016/j.physletb.2011.05.046
[arXiv:1104.2747 [hep-ex]].


\bibitem{Andreas:2012mt}
S.~Andreas, C.~Niebuhr and A.~Ringwald,
Phys. Rev. D \textbf{86}, 095019 (2012)
doi:10.1103/PhysRevD.86.095019
[arXiv:1209.6083 [hep-ph]].

\bibitem{Bjorken:2009mm}
J.~D.~Bjorken, R.~Essig, P.~Schuster and N.~Toro,
Phys. Rev. D \textbf{80}, 075018 (2009)
doi:10.1103/PhysRevD.80.075018
[arXiv:0906.0580 [hep-ph]].

\bibitem{Endo:2012hp}
M.~Endo, K.~Hamaguchi and G.~Mishima,
Phys. Rev. D \textbf{86}, 095029 (2012)
doi:10.1103/PhysRevD.86.095029
[arXiv:1209.2558 [hep-ph]].

\bibitem{Lees:2014xha}
J.~P.~Lees \textit{et al.} [BaBar],
Phys. Rev. Lett. \textbf{113}, no.20, 201801 (2014)
doi:10.1103/PhysRevLett.113.201801
[arXiv:1406.2980 [hep-ex]].

\bibitem{Batley:2015lha}
J.~R.~Batley \textit{et al.} [NA48/2],
Phys. Lett. B \textbf{746}, 178-185 (2015)
doi:10.1016/j.physletb.2015.04.068
[arXiv:1504.00607 [hep-ex]].

\bibitem{Banerjee:2018vgk}
D.~Banerjee \textit{et al.} [NA64],
Phys. Rev. Lett. \textbf{120}, no.23, 231802 (2018)
doi:10.1103/PhysRevLett.120.231802
[arXiv:1803.07748 [hep-ex]].

\bibitem{Banerjee:2019hmi}
D.~Banerjee \textit{et al.} [NA64],
Phys. Rev. D \textbf{101}, no.7, 071101 (2020)
doi:10.1103/PhysRevD.101.071101
[arXiv:1912.11389 [hep-ex]].

\bibitem{LHCb:2017trq}
R.~Aaij \textit{et al.} [LHCb],
Phys. Rev. Lett. \textbf{120}, no.6, 061801 (2018)
doi:10.1103/PhysRevLett.120.061801
[arXiv:1710.02867 [hep-ex]].

\bibitem{LHCb:2019vmc}
R.~Aaij \textit{et al.} [LHCb],
Phys. Rev. Lett. \textbf{124}, no.4, 041801 (2020)
doi:10.1103/PhysRevLett.124.041801
[arXiv:1910.06926 [hep-ex]].

\bibitem{Ilten:2018crw}
P.~Ilten, Y.~Soreq, M.~Williams and W.~Xue,
JHEP \textbf{06}, 004 (2018)
doi:10.1007/JHEP06(2018)004
[arXiv:1801.04847 [hep-ph]].

\bibitem{Chang:2016ntp}
J.~H.~Chang, R.~Essig and S.~D.~McDermott,
JHEP \textbf{01}, 107 (2017)
doi:10.1007/JHEP01(2017)107
[arXiv:1611.03864 [hep-ph]].

\bibitem{deSalas:2016ztq}
P.~F.~de Salas and S.~Pastor,
JCAP \textbf{07}, 051 (2016)
doi:10.1088/1475-7516/2016/07/051
[arXiv:1606.06986 [hep-ph]].

\bibitem{Bashinsky:2003tk}
S.~Bashinsky and U.~Seljak,
Phys. Rev. D \textbf{69}, 083002 (2004)
doi:10.1103/PhysRevD.69.083002
[arXiv:astro-ph/0310198 [astro-ph]].

\bibitem{Hou:2011ec}
Z.~Hou, R.~Keisler, L.~Knox, M.~Millea and C.~Reichardt,
Phys. Rev. D \textbf{87}, 083008 (2013)
doi:10.1103/PhysRevD.87.083008
[arXiv:1104.2333 [astro-ph.CO]].

\bibitem{Baumann:2015rya}
D.~Baumann, D.~Green, J.~Meyers and B.~Wallisch,
JCAP \textbf{01}, 007 (2016)
doi:10.1088/1475-7516/2016/01/007
[arXiv:1508.06342 [astro-ph.CO]].

\bibitem{Friedland:2007vv}
A.~Friedland, K.~M.~Zurek and S.~Bashinsky,
[arXiv:0704.3271 [astro-ph]].

\bibitem{Follin:2015hya}
B.~Follin, L.~Knox, M.~Millea and Z.~Pan,
Phys. Rev. Lett. \textbf{115}, no.9, 091301 (2015)
doi:10.1103/PhysRevLett.115.091301
[arXiv:1503.07863 [astro-ph.CO]].

\bibitem{Blinov:2020hmc}
N.~Blinov and G.~Marques-Tavares,
JCAP \textbf{09}, 029 (2020)
doi:10.1088/1475-7516/2020/09/029
[arXiv:2003.08387 [astro-ph.CO]].

\bibitem{Berlin:2017ftj}
A.~Berlin and N.~Blinov,
Phys. Rev. Lett. \textbf{120}, no.2, 021801 (2018)
doi:10.1103/PhysRevLett.120.021801
[arXiv:1706.07046 [hep-ph]].

\bibitem{Chacko:2003dt}
Z.~Chacko, L.~J.~Hall, T.~Okui and S.~J.~Oliver,
Phys. Rev. D \textbf{70}, 085008 (2004)
doi:10.1103/PhysRevD.70.085008
[arXiv:hep-ph/0312267 [hep-ph]].

\bibitem{Chacko:2004cz}
Z.~Chacko, L.~J.~Hall, S.~J.~Oliver and M.~Perelstein,
Phys. Rev. Lett. \textbf{94}, 111801 (2005)
doi:10.1103/PhysRevLett.94.111801
[arXiv:hep-ph/0405067 [hep-ph]].

\bibitem{EscuderoAbenza:2020cmq}
M.~Escudero Abenza,
JCAP \textbf{05}, 048 (2020)
doi:10.1088/1475-7516/2020/05/048
[arXiv:2001.04466 [hep-ph]].

\bibitem{Hu:2000ti}
W.~Hu, M.~Fukugita, M.~Zaldarriaga and M.~Tegmark,
Astrophys. J. \textbf{549}, 669 (2001)
doi:10.1086/319449
[arXiv:astro-ph/0006436 [astro-ph]].

\bibitem{Hou:2011ec}
Z.~Hou, R.~Keisler, L.~Knox, M.~Millea and C.~Reichardt,
Phys. Rev. D \textbf{87}, 083008 (2013)
doi:10.1103/PhysRevD.87.083008
[arXiv:1104.2333 [astro-ph.CO]].

\bibitem{Hindmarsh:2005ix}
M.~Hindmarsh and O.~Philipsen,
Phys. Rev. D \textbf{71}, 087302 (2005)
doi:10.1103/PhysRevD.71.087302
[arXiv:hep-ph/0501232 [hep-ph]].


\end{thebibliography}
\end{document}